\documentclass[twocolumn]{aastex631}

\usepackage{amsmath}
\usepackage{multirow}

\newcommand{\fb}{f_{\rm{b}}}
\newcommand{\fbhigh}{f_{\rm{b},15}}
\newcommand{\fbhighmod}{f'_{\rm{b},15}}
\newcommand{\qhigh}{q_{15}}
\newcommand{\qbh}{q_{\rm{BH}}}
\newcommand{\Nhigh}{N_{15}}

\newcommand{\msun}{M_{\odot}}

\newcommand{\zsun}{Z_{\odot}}
\newcommand{\pc}{\rm{pc}}

\newcommand{\mcl}{M_{\rm{cl}}}

\newcommand{\myr}{\rm{Myr}}
\newcommand{\gyr}{\rm{Gyr}}
\newcommand{\kms}{\rm{km\,s^{-1}}}

\newcommand{\rc}{r_{\rm{c}}}
\newcommand{\rv}{r_{\rm{v}}}

\newcommand{\cmc}{\texttt{CMC}}
\newcommand{\cosmic}{\texttt{COSMIC}}

\newcommand{\tseg}{t_{\rm{seg}}}
\newcommand{\tms}{t_{\rm{MS}}}

\newcommand{\tcluster}{t_{\rm{cl}}}
\newcommand{\tmerge}{t_{\rm{merge}}}
\newcommand{\mchirp}{M_{\rm{chirp}}}
\newcommand{\mprimary}{M_{\rm{prim}}}
\newcommand{\msecondary}{M_{\rm{sec}}}
\newcommand{\ecc}{{e}}

\newcommand{\Rstar}{{R_\star}}
\newcommand{\Mstar}{{M_\star}}
\newcommand{\nstar}{{n_\star}}

\newcommand{\mbh}{{M_{\rm{BH}}}}
\newcommand{\mbhmax}{M_{\rm{BH,max}}}

\shorttitle{Role of high-mass-star binaries in creating high-mass BHs}
\shortauthors{Khurana \& Chatterjee}

\begin{document}

\title{The Role of High-mass Stellar Binaries in the Formation of High-mass Black Holes in Dense Star Clusters}

\author[0000-0003-3259-6702]{Ambreesh Khurana}
\affil{Department of Astronomy \& Astrophysics, Tata Institute of Fundamental Research, Homi Bhabha Road, Navy Nagar, Colaba, Mumbai 400005, India}
\email{ambreesh.sci@gmail.com}

\author[0000-0002-3680-2684]{Sourav Chatterjee}
\affil{Department of Astronomy \& Astrophysics, Tata Institute of Fundamental Research, Homi Bhabha Road, Navy Nagar, Colaba, Mumbai 400005, India}
\email{souravchatterjee.tifr@gmail.com}

\defcitealias{Gonzalez2021}{GKC21}

\begin{abstract}

Recent detections of gravitational waves from mergers of binary black holes (BBHs) with pre-merger source-frame individual masses in the so-called upper mass-gap, expected due to (pulsational) pair instability supernova ((P)PISN), have created immense interest in the astrophysical production of high-mass black holes (BHs). Previous studies show that high-mass BHs may be produced via repeated BBH mergers inside dense star clusters. Alternatively, inside dense star clusters, stars with unusually low core-to-envelope mass ratios can form via mergers of high-mass stars, which then can avoid (P)PISN, but produce high-mass BHs via mass fallback. We simulate detailed star-by-star multi-physics models of dense star clusters using the Monte Carlo cluster evolution code, $\cmc$, to investigate the role of primordial binary fraction among high-mass stars ($\geq15\,\msun$) on the formation of high-mass BHs. We vary the high-mass stellar binary fraction ($\fbhighmod$) while keeping all other initial properties, including the population of high-mass stars, unchanged. We find that the number of high-mass BHs, as well as the mass of the most massive BH formed via stellar core-collapse are proportional to $\fbhighmod$. In contrast, there is no correlation between $\fbhighmod$ and the number of high-mass BHs formed via BH-BH mergers. Since the total production of high-mass BHs is dominated by BH-BH mergers in old clusters, the overall number of high-mass BHs produced over the typical lifetime of globular clusters is insensitive to $\fbhighmod$. We study the differences in the demographics of BH-BH mergers and their implications for the LIGO-Virgo-Kagra detections as a function of $\fbhighmod$. 

\end{abstract}

\keywords{}

\section{Introduction} 
\label{sec:intro}

Isolated high-mass stars with main-sequence (MS) mass $\gtrsim100\,\msun$ (He core mass $\gtrsim40\,\msun$), during their final stages of evolution, can reach core temperature and pressure high enough to initiate electron-positron pair production \citep[e.g.,][]{Rakavy_1967,Heger_2002}. This leads to a sudden decrease in pressure and the stellar core collapses. The resulting thermonuclear explosion(s) due to oxygen burning can either cause the star to lose mass in episodes, resulting in the so-called pulsational pair instability supernova (PPISN) before collapsing to a black hole (BH) whose mass is thus limited, or destroy the star completely, leaving behind no remnant at all as a result of the so-called pair instability supernova \citep[PISN; e.g.,][]{Barkat1967,Woosley2002}.
One the other hand, models suggest, very high-mass stars with an initial mass $\gtrsim 260\,\msun$ (He core mass $\gtrsim 130\,\msun$) can again collapse to create BHs with mass $\mbh\gtrsim 100\,\msun$ due to photo-disintegration in the core \citep[e.g.,][]{Heger_2003,Woosley2002}.
These processes are thus expected to create a gap between $\sim40$ and $\sim100\,\msun$ in the birth mass function of BHs 
\citep[e.g.,][]{Heger_2002}, although, the exact boundaries of this so-called ``upper mass-gap" are still under investigation \citep[e.g.,][]{Farmer2019,Farmer2020,Costa2021,Vink_2021,Farag2022}. 

Interestingly, gravitational-wave (GW) observatories have detected binary BH (BBH) merger events, such as GW190521 \citep{PhysRevLett.125.101102}, GW190403\_051519, GW190426\_190642 \citep{PhysRevD.109.022001}, GW200220\_061928 \citep{PhysRevX.13.041039}, and GW200129\_114245 \citep{Nitz2023}, with individual pre-merger source-frame masses well within the upper mass-gap. These detections have created intense interest in exploring the physical processes that could lead to the formation of such
high-mass BHs. Inside dense star clusters, BH-BH binaries are assembled via dynamical processes which then can merge via the emission of GWs inside the cluster or after getting ejected from the cluster \citep[e.g.,][]{Banerjee_2010,Rodriguez_2015,Rodriguez_2016,Askar_2017,Banerjee_2017,Chatterjee_2017b,Chatterjee_2017,Banerjee_2018,Banerjee_2018b,Samsing_2018,Kremer_2020}. 
If the merger product is not ejected from the cluster, the newly formed BH from the BBH merger may acquire another BH companion via the same dynamical processes and merge again \citep[e.g.,][]{Rodriguez_2018,Rodriguez_2020c,Martinez_2020}. This process can continue until the merger product is ejected from the cluster either via recoil from dynamical interactions or merger-driven kicks. It has been proposed that BH-BH mergers involving BHs within the upper mass gap may have been formed via these so-called higher generation mergers \citep[e.g.,][]{PhysRevD.100.043027,Weatherford2021}. 

On the other hand, inside dense star clusters, at early times, high-mass stars may collide with other stars. These collision products can have a high enough mass to form a mass-gap BH but have unusually low core-to-envelope mass ratios \citep[e.g.,][]{Ballone_2022} such that the core mass is not high enough to initiate (P)PISN. Thus, the star does not lose too much mass during core-collapse\footnote{The phrase `core-collapse' can be used in different contexts- the collapse of a star's core before remnant formation and the collapse of a cluster's core due to two-body relaxation and mass segregation. Unfortunately, in this paper both of these could be applicable. To avoid confusion, we use this phrase exclusively to mean stellar core-collapse throughout the paper.}. This could create a BH more massive than what is possible via single star evolution as the initial BH could grow in mass via at least partial fallback of the unusually massive envelope \citep[e.g.,][]{DiCarlo2019,Spera_2019,DiCarlo2020,Kremer2020b,Renzo2020,Gonzalez2021,Costa_2022,2024arXiv240505397S}. Note that both these processes, i.e., BH-BH mergers and core-collapse of stellar merger products, may undergo within dense star clusters and the two processes are not necessarily independent \citep[e.g.,][]{Sedda2023,Prieto_2022}. For example, a high-mass BH, formed via the latter mechanism, may later take part in multiple BH-BH mergers via the former pathway. It is interesting, however, to understand the relative importance of these processes as a function of the initial cluster properties and age. 

Theoretical models and observations suggest that high-mass stars are often in binaries or higher-order multiples \citep[e.g.,][]{Sana2012,Offner2023}. Recently, \citet{Gonzalez2021} (hereafter \citetalias{Gonzalez2021}) showed that massive star clusters, modeled using the {\tt Cluster\ Monte\ Carlo} (\cmc) code \citep[][]{2000ApJ...540..969J,2001ApJ...550..691J,2007ApJ...658.1047F,2010ApJ...719..915C,2012ApJ...752...43G,Morscher_2013,2013ApJS..204...15P,Breivik2020,Rodriguez2022}, 
with an initial binary fraction among high-mass ($>15\,\msun$) stars, $\fbhigh=1$, can produce BHs with mass in the upper mass-gap and even higher (collectively, ``high-mass BHs", hereafter) via core-collapse of stellar merger products. In contrast, models with $\fbhigh=0$, produced almost no high-mass BHs \citepalias[][in particular, see their Figure\ 2]{Gonzalez2021}.\footnote{$\fbhigh \equiv N_{b,15}/(N_{b,15}+N_{s,15})$, where $N_{b,15}$ and $N_{s,15}$ represent the number of binaries containing at least one high-mass star and the number of high-mass single stars, respectively. Similarly, $f_b\equiv N_b/(N_b+N_s)$, where $N_b$ ($N_s$) denotes the number of low-mass binaries (single).}  There were two important differences between the two sets of models with $\fbhigh=0$ and $1$. First, all high-mass stars in $\fbhigh=1(0)$ models were in binaries (singles). The other, relatively less apparent, difference is in the total initial number of high-mass stars. In \cmc, the masses of stars, primaries in case of binaries ($\mprimary$), are sampled assuming an initial stellar mass function (IMF). Then, depending on the assumed overall and high-mass binary fractions, $\fb$ and $\fbhigh$, a randomly chosen fraction of these stars (or primaries) are assigned a binary companion with mass ($\msecondary$) based on the adopted mass-ratio ($q\equiv\msecondary/\mprimary$) distribution and range. Thus, for a fraction of the companions, $\msecondary>15\,\msun$. As a result, models with $\fbhigh=1$ end up with a higher number of high-mass stars than the models with $\fbhigh=0$. In other words, changing $\fbhigh$ does not conserve the overall IMF. For example, using the adopted initial conditions in \citetalias{Gonzalez2021}, i.e., $N=8\times10^5$ (number of initial objects, singles and binaries), stellar IMF based on \citet{Kroupa2001}, uniform mass ratio distribution for high-mass stars between $\qhigh=0.6$ and $1$, and $\fbhigh=1$, the total number of initial stars with mass $\geq15\,\msun$, $\Nhigh=3782^{+130}_{-63}$. In contrast, using the same assumptions, but with $\fbhigh=0$, $\Nhigh=2181^{+74}_{-32}$. This leads to a difference in initial $\Nhigh$ of $1606^{+53}_{-44}$. Interestingly, \citetalias{Gonzalez2021} also found that most ($\approx 96\%$) high-mass BHs formed via stellar collisions during dynamical encounters as opposed to binary stellar evolution.  
Thus, it is not clear whether the formation of much higher numbers of high-mass BHs in their $\fbhigh=1$ models is because of the difference in $\fbhigh$ or $\Nhigh$. Moreover, given the importance of high-mass stars in the overall evolution and even survival of the host cluster and the formation of stellar exotica \citep[e.g.,][]{Chatterjee_2017} 
disentangling the effects of the differences in $\Nhigh$ and $\fbhigh$ is very interesting.  

In this paper, 
we investigate whether the high-mass BHs that form in star cluster models with $\fbhigh=1$ indeed require those high-mass stars to be in a binary initially or this is simply a result of a higher $\Nhigh$ in these models compared to those with $\fbhigh=0$. We create models systematically varying the initial fraction of high-mass stars in binaries between $0$ and $1$ keeping $\Nhigh$ and other initial properties fixed.   
We investigate the long-term evolution of these model clusters paying particular attention to the formation of high-mass BHs via core-collapse of stellar merger products as well as BH-BH mergers. Furthermore, we investigate the relative contribution to high-mass BH-BH mergers from these two channels as a function of the cluster age. 

In \autoref{sec:methods}, we describe our star cluster models. We also describe the process by which we initialise cluster models with varying $\fbhigh$ keeping $\Nhigh$ and other properties unchanged. In \autoref{sec:results}, we present our findings in detail. Finally, we conclude and provide a summary in \autoref{sec:conclusions}.

\section{Methods}
\label{sec:methods}

We use the highly realistic Hénon-type Monte Carlo cluster evolution code, $\cmc$ \citep{Rodriguez2022} to simulate star clusters. The stellar evolution in $\cmc$ is handled by the population synthesis code, $\cosmic$ \citep{Breivik2020}. As the so-called `control group' of our cluster models, we choose the initial parameters as follows. This group of models has initial $\fbhigh=1$ and the binary fraction in lower-mass stars, $\fb=0.05$. The initial $N=8\times10^5$. The single and primary stars are sampled from a \citet{Kroupa2001} IMF ranging from $0.08$ to $150\,\msun$. For the binaries, eccentricities ($\ecc$) are assumed to be thermal \citep[$dN/d\ecc=2\ecc,$][]{Jeans1919} while the orbital periods ($P$) are drawn from the distribution, $dN/d\log{P}\propto P^{-0.55}$ \citep{Sana2012} between $\log{(P/\rm{day})}=0.15$ and the smaller of the hard-soft boundary and $\log{(P/\rm{day})}=5.5$. The mass ratio between binary members is sampled from a uniform distribution between $q=0.1$ and $1$ for lower-mass primaries and $\qhigh=0.6$ and $1$ for the high-mass primaries. This results in a total initial cluster mass $\mcl/\msun\sim5.4\times10^5$. The cluster members are distributed according to the \citet{King1966} profile with a concentration parameter, $w_0=5$ and a virial radius, $\rv=1\,\pc$. The metallicity of the stars is fixed at $0.1\,\zsun$, where $\zsun=0.02$. We simulate each cluster up to $\approx13.8\,\gyr$. In order to take statistical fluctuations into account, we run 10 models with the same properties but generated using different random seeds. 

To explore the importance of initial $\fbhigh$ compared to $\Nhigh$ in creating high-mass BHs, we create additional cluster models which have the same initial $\Nhigh$ as the control group but different $\fbhigh$. For each model in our control group with $\fbhigh=1$, we randomly choose $25\%$, $50\%$, $75\%$, and {\em all} high-mass stellar binaries and break them into singles. We redistribute the new collection of single and binary stars in the cluster according to the same \citet{King1966} profile. This leaves us with a total number of stellar objects (binary/single) higher than $8\times10^5$. This procedure helps us create sets of initial cluster models with $\fbhighmod=0.75$, $0.5$, $0.25$, and $0$, respectively, where the individual stellar populations remain identical between corresponding cluster models. Here, $\fbhighmod$ denotes the fraction of initially high-mass binaries that we leave intact. The changed notation indicates that $\fbhighmod$ and $\fbhigh$ are not equal unless $\fbhighmod=0$ or $1$ (\autoref{tab:ccbh}). If the production of high-mass BHs requires enough number of high-mass stars in the cluster independent of whether they are initially in a binary or not, then these cluster models with different initial $\fbhighmod$ but the same $\Nhigh$ would show little statistical difference. On the other hand, if $\fbhighmod$ is indeed a critical property, then we would find systematic differences between these groups of models.

The formation of high-mass BHs via core-collapse of stellar merger products depends on how stellar collisions are treated in the models. \citetalias{Gonzalez2021} used the simple `sticky sphere' approximation for MS and giant star collisions \citep[section 2, ][]{Kremer2020b}. In contrast, we use the updated prescriptions for collisions involving giant stars currently used in $\cmc$ \citep{Ye2022a}. These updates are motivated by the expectation that collisions involving giant stars should be better represented by something similar to a common envelope (CE) evolution in which the core of the giant and the companion (a MS star or a compact object) spiral towards each other losing orbital energy to the giant's envelope that enshrouds them. 
If the envelope can be ejected due to the energy released by the shrinking orbit before they merge, a binary should form composed of the giant's core and the secondary object. If both of the colliding stars are giants, the binary formed would be composed of the cores of the two giants. However, if the envelope can not be ejected, they merge. The updated prescription calculates orbital parameters for the final binary assuming a CE evolution \citep[][]{Hurley2002}
as well as from a set of parameterised equations \citep{Ivanova2006} adapted from the results of smoothed particle hydrodynamics simulations \citep{LombardiJr.2006}. If not merged, the final binary is assigned the orbital properties corresponding to the smaller semi-major axis ($a$) from the above two solutions. If the formed binary satisfies the condition for Roche-lobe overflow, the participating stars are merged including the envelope \citep[for more details on the updated prescription, see][]{Ye2022a}.

In addition to the treatment of stellar collisions, our results also depend on the adopted prescriptions for modeling BH natal kicks and BH-BH merger recoil because these affect the retention of BHs in the cluster after they form and after a GW-driven merger, respectively.  
We assume that BHs receive natal kicks scaled down by the mass fallback fraction \citep{Fryer_2012} from the standard neutron star (NS) kicks, where NS natal kicks are drawn from a Maxwellian distribution with dispersion $265\,\kms$ \citep{10.1111/j.1365-2966.2005.09087.x,Kremer_2020}. We also assume that BHs are born non-spinning \citep[][]{Breivik2020}. The mass, spin, and GW recoil of the GW-driven merger products are estimated using the prescriptions given in \citet{Gerosa2016} based on numerical relativity calculations.\footnote{See \citet{Rodriguez_2018} Appendix A for more details and a typo correction in the remnant-mass expression.} 

In our simulations, we choose the prescription for (P)PISN physics parameterised from the simulations of \citet{Marchant2019}, available by default in $\cmc$-$\cosmic$ \citep{Breivik2020}. This corresponds to an upper mass-gap in the range $\approx 44.3-123\,\msun$. Furthermore, we use the ``delayed" supernova (SN) explosion model for determining the mass of the remnant after SN \citep{Fryer_2012}. 

\section{Results}
\label{sec:results}
\begin{figure}
    \centering
    \epsscale{1.2}
    \plotone{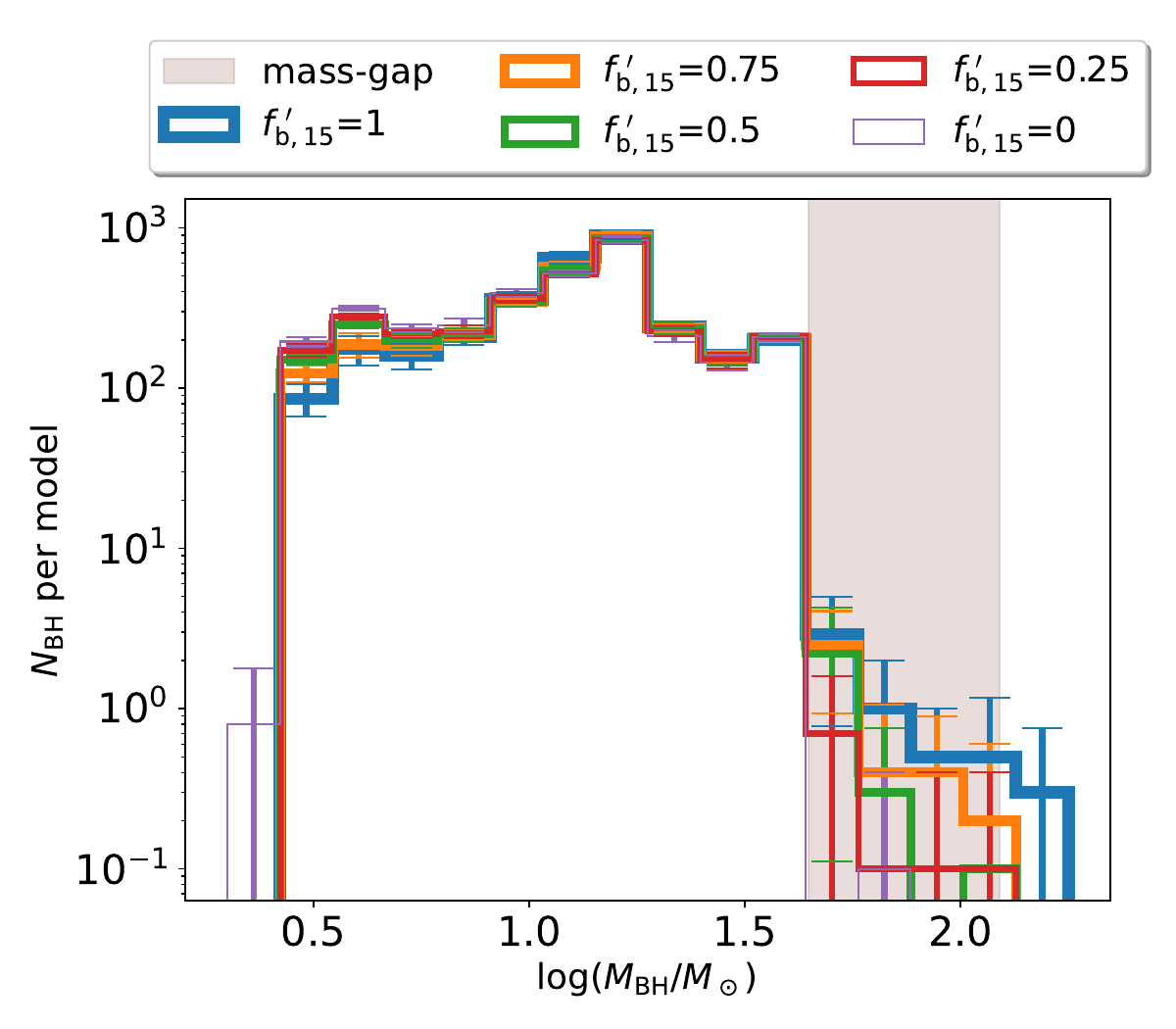}
    \caption{Distribution of the average number of BHs produced via stellar core-collapse per model as a function of BH mass. Errorbars represent 1$\sigma$. The vertical shaded region represents the upper mass-gap due to (P)PISN. Blue, orange, green, red, and purple denote models with $\fbhighmod=1$, $0.75$, $0.5$, $0.25$, and $0$, respectively. Although, the $\mbh$ distributions in cluster models with different $\fbhighmod$ are similar below the upper mass-gap, the number of high-mass BHs produced depends on $\fbhighmod$.}
    \label{fig:mbh_hist}
\end{figure}
\begin{figure}
    \centering
    \epsscale{1.1}
    \plotone{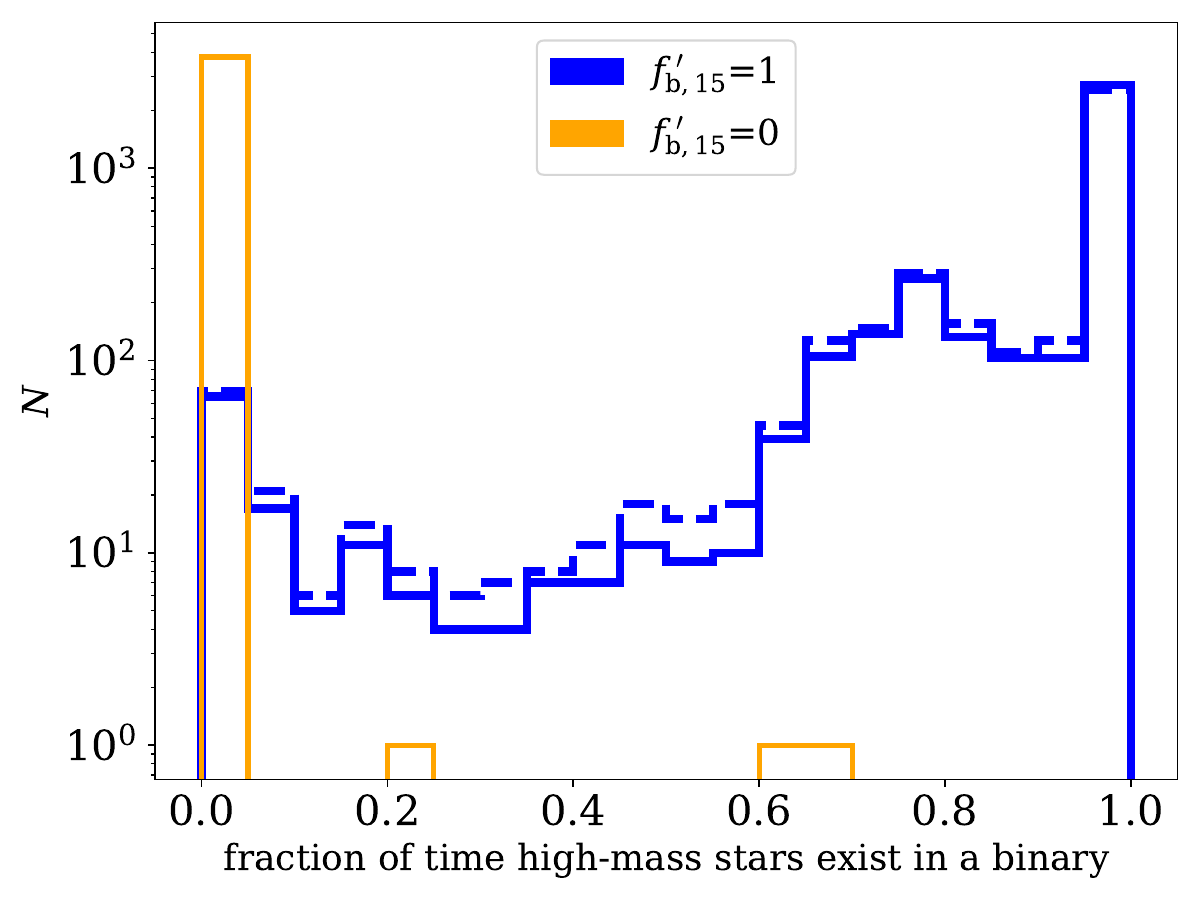}
    \caption{Fraction of time the high-mass ($\Mstar/\msun>15$) stars exist as a part of a binary during the first $35\,\myr$, before either collapsing to a compact object, colliding with another star, or getting ejected from the cluster. Blue and orange denote example models with $\fbhighmod=1$ and $0$. Solid and dashed denote time fractions spent as a binary overall, and with the primordial companion, respectively.}
    \label{fig:time_frac}
\end{figure}
\begin{figure}
    \centering
    \epsscale{1.1}
    \plotone{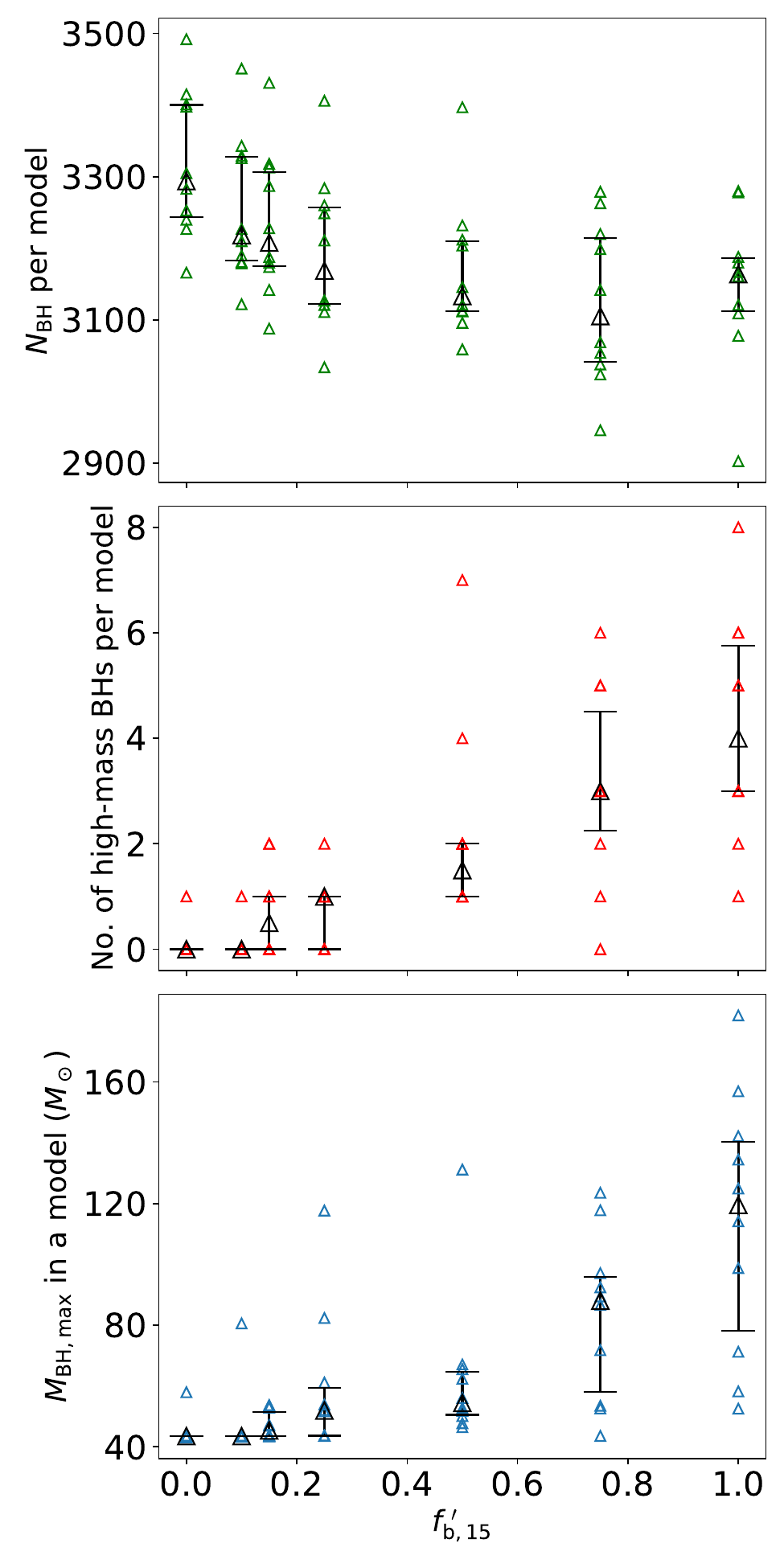}
    \caption{The total number of BHs (\emph{top}), high-mass BHs (\emph{middle}), and the mass of the most massive BH (\emph{bottom}) formed due to stellar core-collapse in a cluster model as a function of $\fbhighmod$. Colored triangles represent individual realisations for the model clusters. Black triangles with errorbars denote the median and the quartiles for all model realisations with a given $\fbhighmod$.}
    \label{fig:nbh}
\end{figure}

\begin{figure*}
    \centering
    \epsscale{1.15}
    \plotone{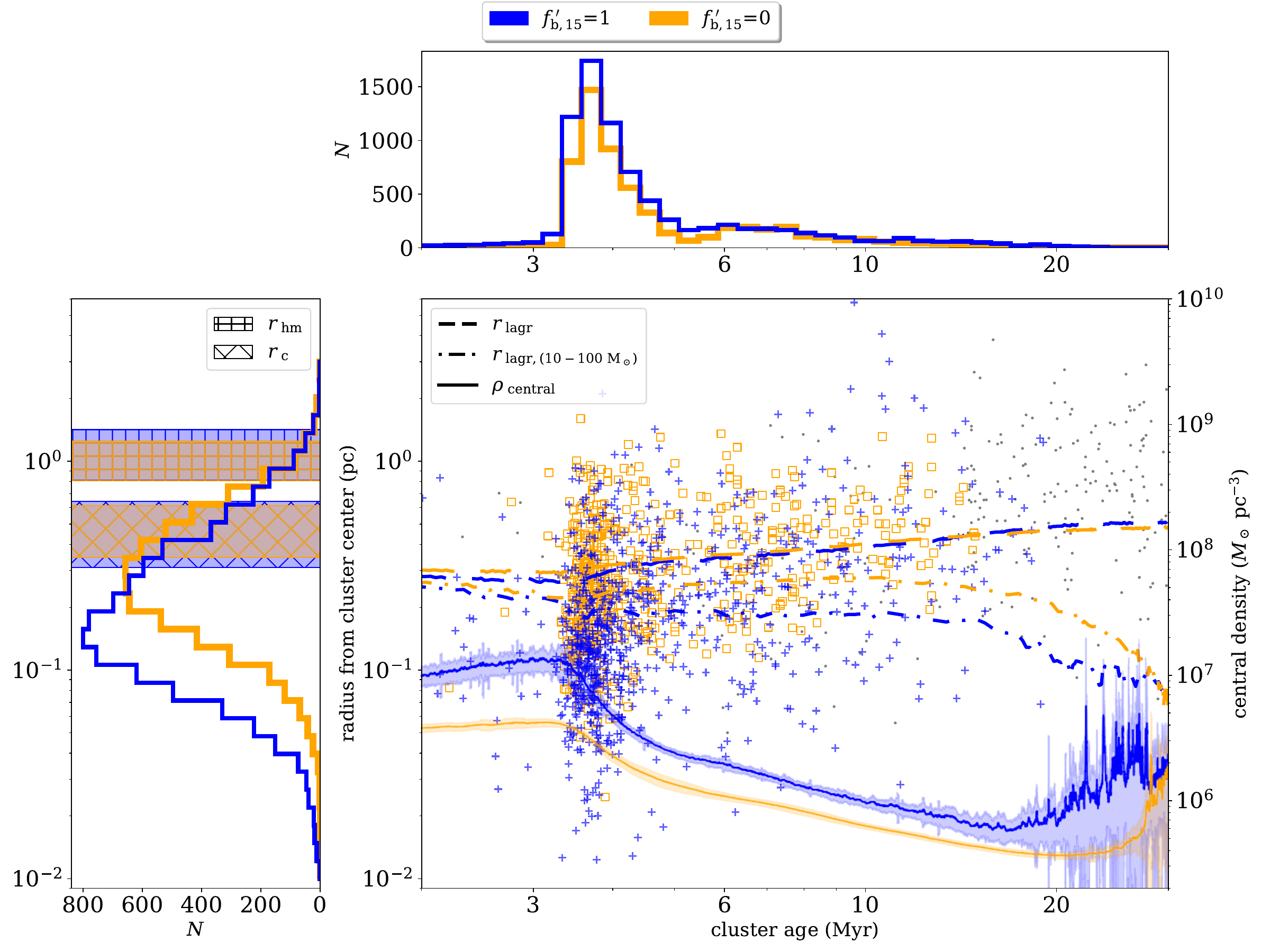}
    \caption{All stellar collisions/mergers from example cluster models. In the main panel, we show examples from single realisations of models with $\fbhighmod=1$ (blue) and $0$ (orange). The colored markers and grey dots denote all collisions/mergers involving at least one high-mass star and those without any high-mass star involvement, respectively. We show the $10\%$ Lagrange radii for all stars (dashed) and stars with mass $10<M_\star/\msun<100$ (dash-dot). We also show the average central density $\rho_c$ (solid with $1\sigma$ range denoted by the shaded regions), the scale for $\rho_c$ is shown on the right vertical axis. The top and left panels show the corresponding distributions for high-mass stellar collision/merger times and locations from all models with $\fbhighmod=1$ (blue) and $0$ (orange), respectively. The hatched-shaded regions in the left panel represent the ranges of the half-mass radius (`+'-hatched) and the core radius (`x'-hatched) the cluster models attain during this early evolution.}
    \label{fig:coll_time}
\end{figure*}

\begin{figure}
    \centering
    \epsscale{1.2}
    \plotone{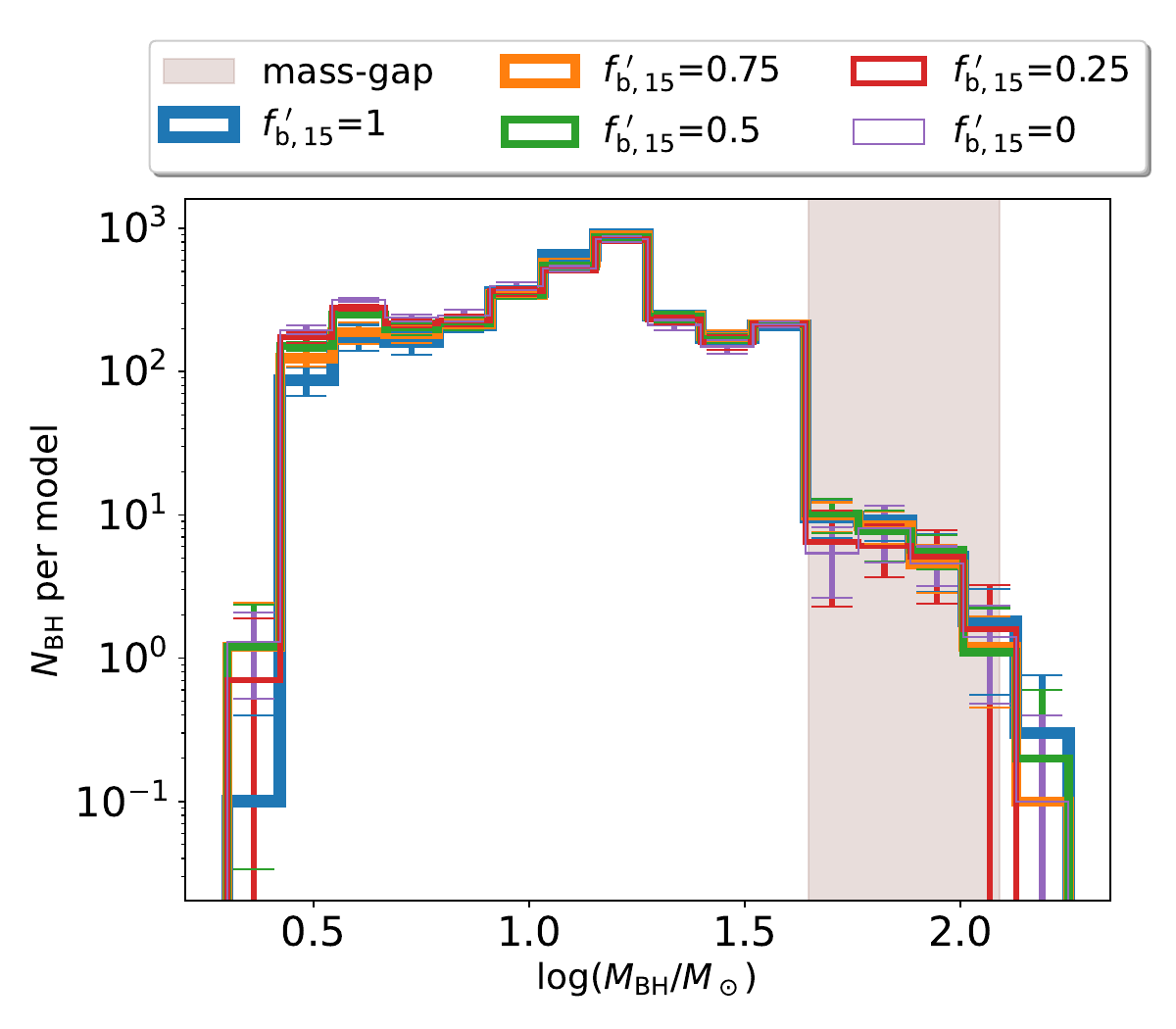}
    \caption{The same as \autoref{fig:mbh_hist}, except here we also consider the BHs that form via BH-BH mergers throughout the Hubble time of evolution.}
    \label{fig:mbh_hist2}
\end{figure}

\begin{figure}
    \centering
    \epsscale{1.1}
\plotone{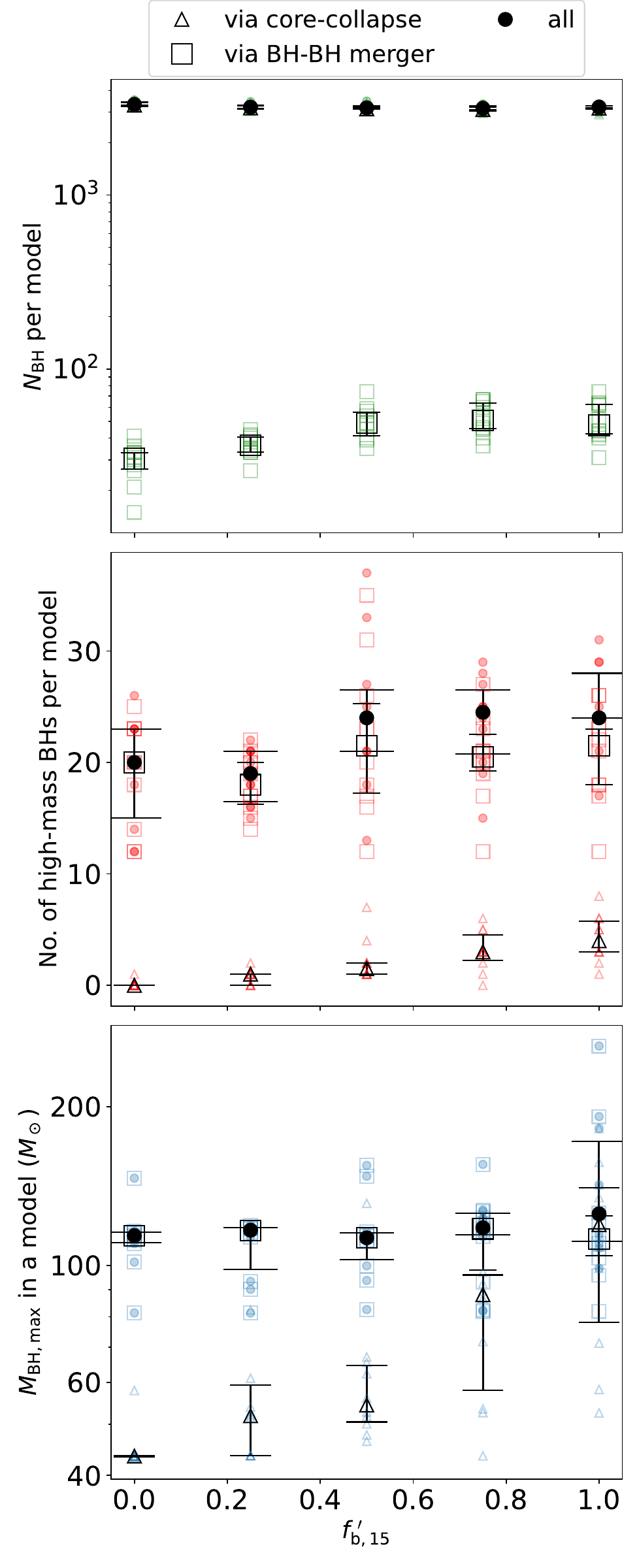}
    \caption{The same as \autoref{fig:nbh}, except here we also consider the BHs that form via BH-BH mergers throughout the Hubble time of evolution. Furthermore, we divide the complete population of BHs (\emph{filled circles}) into two groups: the ones that form via stellar core-collapse (\emph{unfilled triangles}) and the ones that form via BH-BH mergers (\emph{unfilled squares}).
    }
    \label{fig:bhsall}
\end{figure}

\begin{figure*}
    \centering
    \epsscale{1.15}
    \plotone{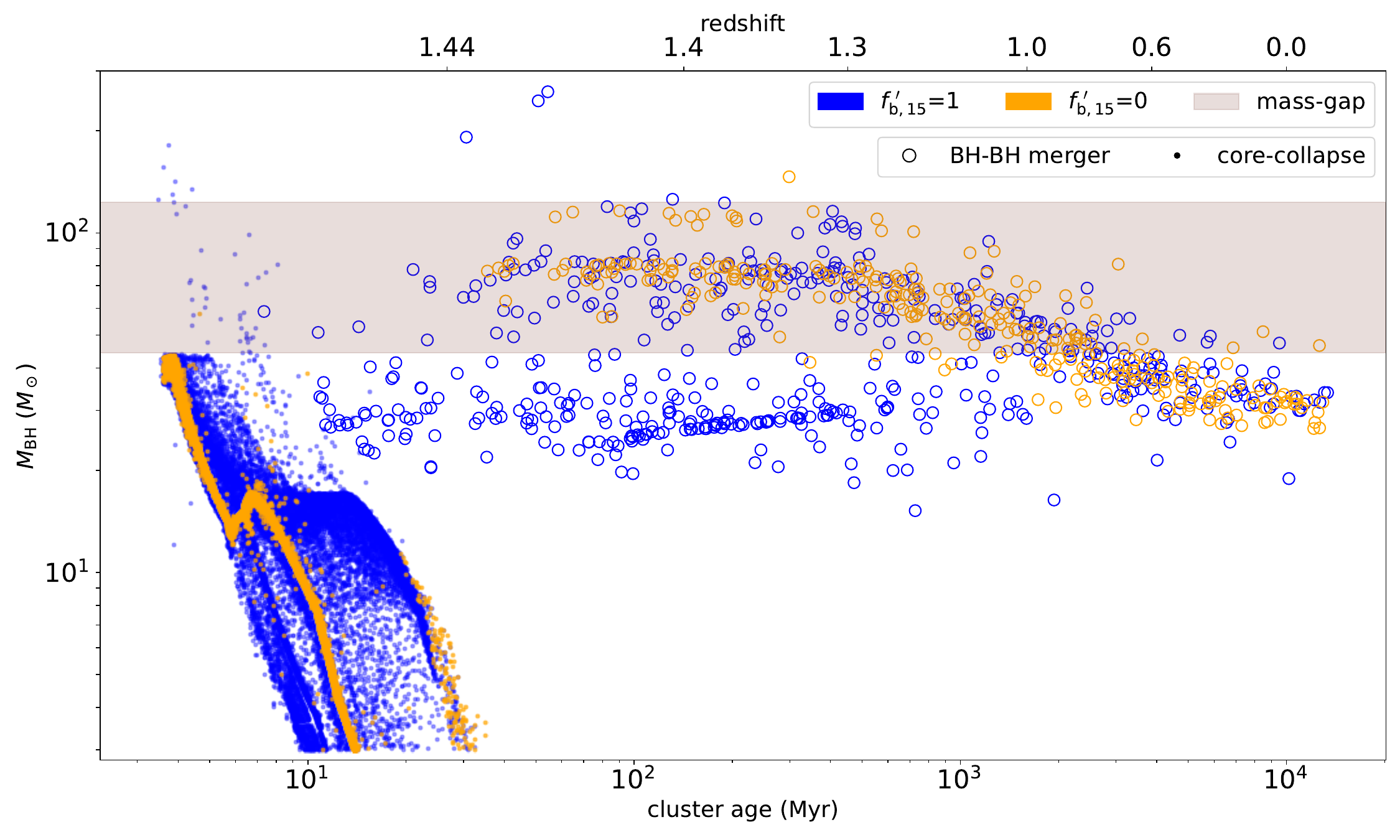}
    \caption{The birth mass and the time of formation of all the BHs in our cluster models. Blue (orange) denotes $\fbhighmod=1$ ($0$). We include all BHs formed in the 10 different realisations of cluster models in both the groups. The shaded region denotes the upper mass-gap. The ``$\circ$" (``$\boldsymbol{\cdot}$") represents BHs formed via BH-BH mergers (stellar core-collapse). The top axis represents indicative redshift, $z$, calculated assuming cosmic time, $t_c=14\,\gyr$ represents $z=0$ \citep{Carmeli2006} and the model star clusters formed at $t_c=4\,\gyr$, i.e., their present age is $10\,\gyr$.}
    \label{fig:mbh_time}
\end{figure*}

\subsection{Evolution in Early Times}
\label{subs:earlytime}

\begin{table*}
    \centering
    \hspace{-2cm}\begin{tabular}{cccccccc}
    \hline
        \multirow{2}{.7cm}{$\fbhighmod$} & \multirow{2}{.7cm}{$\fbhigh$} & \multicolumn{3}{c}{$N_{\rm{BH,high}}$} & \multicolumn{3}{c}{$M_{\rm{BH,max}}/\msun$} \\\cline{3-8}
         & & Core-collapse & BH-BH Merger & Mass Transfer & Core-collapse & BH-BH Merger & Mass Transfer \\
         \hline \hline
        1 & 1 & $4_{-1}^{+1.75}$ & $21.5_{-3.5}^{+1.5}$ & $20_{-5}^{+4}$ & $119.5_{-41.4}^{+20.7}$ & $112.3_{-7.8}^{+11.8}$ & $50.7_{-2.1}^{+1.5}$ \\
        0.75 & 0.63 & $3_{-0.75}^{+1.5}$ & $20.5_{-1.25}^{+2}$ & $15_{-2}^{+3}$ & $88_{-30}^{+7.9}$ & $117.3_{-19.3}^{+8.3}$ & $51.8_{-3.3}^{+2.9}$ \\
        0.5 & 0.36 & $1.5_{-0.5}^{+0.5}$ & $21.5_{-4.25}^{+3.75}$ & $10_{-1}^{+2}$ & $54.4_{-3.9}^{+10.3}$ & $112.8_{-10.2}^{+2.5}$ & $51.8_{-1.0}^{+1.4}$ \\
        0.25 & 0.16 & $1_{-1}^{+0}$ & $18_{-1.75}^{+2}$ & $4_{-1}^{+1}$ & $51.9_{-8.3}^{+7.4}$ & $116.5_{-18.2}^{+1.6}$ & $48.1_{-0.5}^{+0.3}$ \\
        0 & 0 & $0_{-0}^{+0}$ & $20_{-5}^{+3}$ & $0_{-0}^{+0}$ & $43.5_{-0.1}^{+0.1}$ & $113.9_{-3.5}^{+1.6}$ & N/A \\
        \hline
    \end{tabular}
    \caption{Demographics of the BH populations created in our star cluster models with varying $\fbhighmod$. For each $\fbhighmod$ we show the corresponding $\fbhigh$. We show the number of BHs with mass in the upper mass gap or above ($N_{\rm{BH,high}}$) and the maximum BH mass ($\mbhmax$). For each of the above, we separately show BHs created via stellar core-collapse, BH-BH mergers, and mass transfer in a binary. The numbers and errorbars denote the median and the quartiles.}
    \label{tab:ccbh}
\end{table*}

To study high-mass BH formation in dense star clusters via core-collapse of stellar collision products, we first focus on the early ($<35\,\myr$) stages of evolution. This is the time when high-mass stars exist as luminous objects before turning into remnants such as BHs and NSs. \autoref{fig:mbh_hist} shows the histograms of the mass of all BHs ($\mbh$) produced via stellar core-collapse in our models with various initial $\fbhighmod$. We find that all stellar core-collapse BHs form within a cluster age, $\tcluster/\myr<35$ in our models. 
The vertical shaded region represents the upper mass-gap corresponding to the assumptions in our simulations. Clearly, the high-mass end of the BH mass spectrum is sensitive to the initial $\fbhighmod$-- models with higher initial $\fbhighmod$ produce more high-mass BHs via stellar core-collapse despite the same $\Nhigh$ in all models. 

Being part of a binary initially does not ensure that the binary stays intact. Hence, we investigate the fractions of times high-mass stars remain part of a binary during this early ($\tcluster/\myr<35$) evolution before forming remnants, colliding with another star, or getting ejected from the cluster. \autoref{fig:time_frac} shows the distribution for the fraction of time the high-mass stars remain in a binary system. We find that if initially in a binary, the high-mass stars remain part of a binary most of the time. Moreover, most of the time, the companion is the primordial one. In contrast, the high-mass stars in the $\fbhighmod=0$ models rarely capture a companion so early in the evolution of the cluster. 

The collision rate for stars is proportional to the stellar number density ($\nstar$) and the square of the stellar radius ($\Rstar$). Hence, if the mass segregation timescale ($\tseg$) is shorter than the MS lifetime ($\tms$), then the high-mass stars can sink to the high-density core as large objects significantly boosting the rate of physical collisions. In contrast, if $\tseg>\tms$, then the high-mass stellar collisions are suppressed significantly. By the time high-mass stars can sink to the center, they evolve and form remnants. Formation of high-mass BHs via core-collapse of stellar collision products of course requires that the collisions happen before remnant formation \citep[e.g.,][]{DiCarlo2019}. The mass loss during remnant formation also leads to rapid overall expansion of the cluster leading to a sharp decrease in $\nstar$ at the core \citep[e.g.,][]{2010ApJ...719..915C}. If a high-mass star is part of a binary, then $\tseg$ reduces by a factor of $\sim(1+\qhigh)$ thus increasing the chance that the high-mass star sinks to the core before remnant formation. Moreover, for a binary, the strong interaction cross-section is $\propto a^2$, significantly higher than single-star collision cross section. For the typical velocity dispersion inside dense stellar clusters, the dominant resonant type interactions often lead to physical collisions involving the large high-mass stars \citep[e.g.,][]{Fregeau2004}. 

Indeed, we find a clear correlation between the number of high-mass BHs produced via core collapse and $\fbhighmod$. While the models with $\fbhighmod=1$ create $\approx4$ high-mass BHs via stellar core-collapse per cluster, this number decreases almost linearly to $\approx0$ for the models with $\fbhighmod<0.1$ (\autoref{fig:nbh}, middle). Clusters with higher $\fbhighmod$ are expected to bring in more high-mass stars to the core before remnant formation, which leads to a higher chance of high-mass BH production via core-collapse of collision products. We find that the mass of the most massive BH ($\mbhmax$) a cluster produces via stellar core-collapse depends on $\fbhighmod$ as well (\autoref{fig:nbh}, bottom). While models with $\fbhighmod=1$ could produce BHs with mass as high as $\approx180\,\msun$, the heaviest BH in the $\fbhighmod=0$ models is $\approx60\,\msun$. The median $\mbhmax$ decreases from $\approx120\,\msun$ for $\fbhighmod=1$ to $\approx43\,\msun$ for $\fbhighmod=0$. Increasing $\fbhighmod$ decreases $\tseg$. Hence, the mass of the most massive star satisfying $\tseg<\tms$ increases with increasing $\fbhighmod$, thus increasing $\mbhmax$.\footnote{We find that the level of mass segregation in $\cmc$ models is in excellent agreement with direct $N$-body models, at least for this part of the evolution (\autoref{app:dnb}).} 

Interestingly, the higher-$\fbhighmod$ models produce fewer BHs overall, and in particular, fewer low-mass BHs (\autoref{fig:nbh}, top). This is because, multiple high-mass stars, each of which could potentially produce individual BHs via stellar core-collapse, may collide to produce a single high-mass BH in high-$\fbhighmod$ models. 

Since the high-mass BHs are produced via stellar collisions at early times, we study all collisions in detail. \autoref{fig:coll_time} shows the time and cluster-centric distance ($r$) of all stellar collisions (during dynamical encounters) and mergers (due to stellar evolution processes in a binary) in two representative model clusters, one with $\fbhighmod=1$ (blue) and the other with $\fbhighmod=0$ (orange). Corresponding histograms for $\tcluster$ and $r$ show the distributions created using all realisations. Close to $80\%$ ($95\%$) of all collisions take place within the core radius, $\rc$ (half-mass radius, $r_{\rm{hm}}$). This confirms the importance of quick mass segregation for high-mass stellar collisions. Each $\fbhighmod=1$ model produces $800^{+46}_{-46}$ ($100^{+16}_{-25}$) collisions (mergers) involving at least one high-mass star. A little over half ($449^{+23}_{-92}$) of these collisions are binary mediated, i.e., they occur during a binary-single or a binary-binary encounter. In contrast, only $593^{+7}_{-7}$ ($4^{+1}_{-1}$) high-mass stellar collisions (mergers) occur per model for $\fbhighmod=0$. Of these collisions, only $2^{+1}_{-1}$ per model are binary mediated. 

Interestingly, although, high-mass primaries remain bound to their primordial companions most of the time (\autoref{fig:time_frac}) during $\tcluster/\myr\leq35$, only $\approx3\%$ of the high-mass \emph{stellar collisions} involve the primordial companions in the $\fbhighmod=1$ models. In contrast, almost all ($\approx98\%$) high-mass \emph{mergers} involve primordial companions in these models. Of course, by construct, none of the collisions/mergers in the $\fbhighmod=0$ models involve primordial companions. Overall, high-mass stellar collisions/mergers seem to occur at a statistically smaller $r$ in the models with $\fbhighmod=1$ compared to the locations of those in the models with $\fbhighmod=0$. This is because of the more efficient mass segregation in the higher $\fbhighmod$ models, evident from the evolution of the $10\%$ Lagrange radii ($r_{\rm{lagr}}$). Although $r_{\rm{lagr}}$ for all stars are comparable, the model with $\fbhighmod=1$ has a smaller $r_{\rm{lagr}}$ for $10$--$100\,\msun$ stars than that for the model with $\fbhighmod=0$.

Interestingly, most of the high-mass collisions/mergers occur during a short time window, $3\lesssim\tcluster/\myr\lesssim5$ independent of $\fbhighmod$. This is because of an enticing competition between core contraction due to two-body relaxation and rapid expansion due to stellar mass loss via winds and remnant formation in young star clusters. The evolution of the central density ($\rho_c$) helps understand the process clearly (\autoref{fig:coll_time}). Until $\tcluster/\myr\lesssim3$, $\rho_c$ increases due to two-body relaxation and mass segregation. When $\tcluster/\myr\approx3$, the high-mass stars start undergoing SN. Afterwards, the mass-loss during remnant formation leads to rapid expansion of the whole cluster decreasing $\rho_c$ by about an order of magnitude. Hence, during a small time window, high-mass stars attain their largest sizes immediately before SN and $\rho_c$ remains sufficiently high. This leads to a sharp spike in high-mass stellar collisions. While a more efficient mass segregation leads to a higher $\rho_c$ in the higher $\fbhighmod$ models, host to the same stellar populations, all models independent of $\fbhighmod$ exhibit the spike in high-mass stellar collisions/mergers during the same time window. In this context, it is interesting to discern the effects of dynamics from what is expected from isolated binary stellar evolution of the same high-mass binaries. We find that isolated binary stellar evolution does not produce such a spike of stellar collisions/mergers between $3\leq\tcluster/\myr\leq5$. Moreover, the total number of high-mass stellar collisions/mergers is significantly lower if the same binaries are evolved in isolation. See \autoref{app:isolated} for more details.

\subsection{High-mass BHs via BH-BH Mergers}
\label{subs:bhsall}

\begin{figure}
    \centering
    \epsscale{1.1}
    \plotone{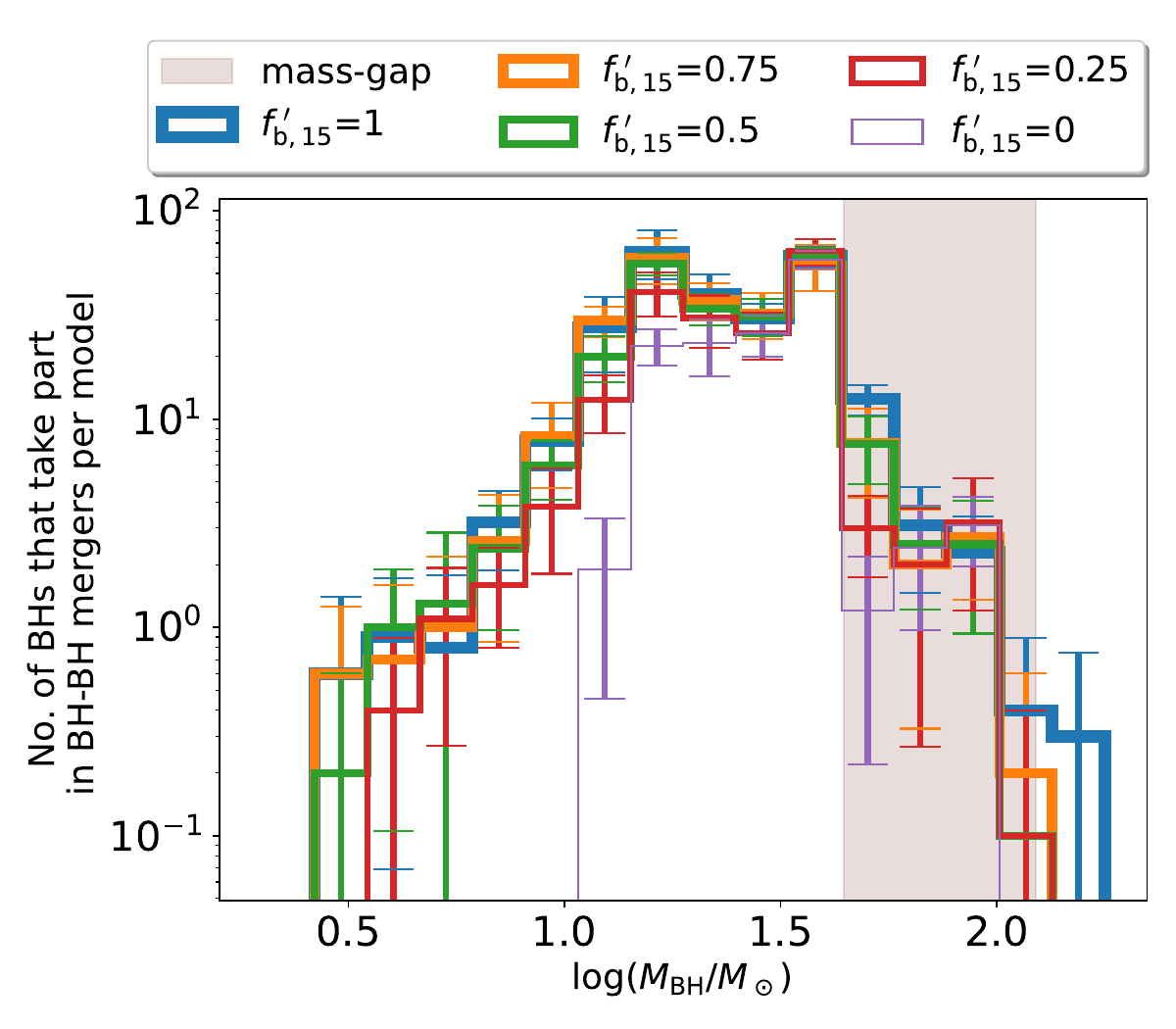}
    \plotone{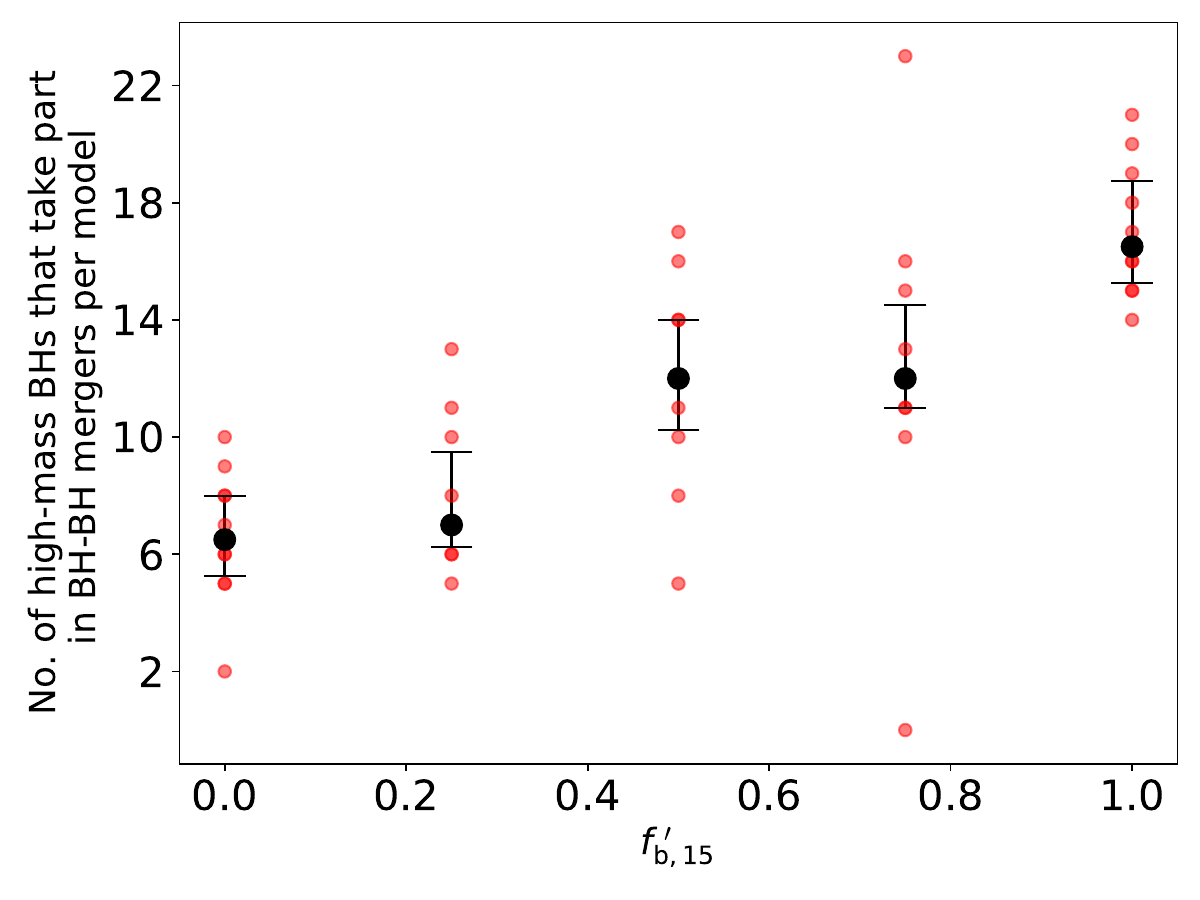}
    \caption{{\em Top}: Distribution of pre-merger masses for merging BBHs from models with different $\fbhighmod$. Colors, errorbars, and shaded region denote the same as in \autoref{fig:mbh_hist}. {\em Bottom}: The number of high-mass merging BHs produced as a function of $\fbhighmod$ over the full simulated time. Colored dots denote individual realisations. Black circles and errorbars denote the median and quartiles for all realisations for a given $\fbhighmod$. 
    }
    \label{fig:bhsmergers}
\end{figure}

So far we have focused on the early life of a star cluster and the formation of high-mass BHs via core-collapse of collision products. However, high-mass BHs can continue to form via BH-BH mergers through dynamical processes until most BHs are ejected from the cluster \citep[e.g.,][]{Rodriguez_2018}.
We now shift our attention to the overall production of all BHs throughout the entire lifetime of the star cluster. In this context, when a BH merges with another BH, we consider that a new BH is formed.\footnote{\cmc\ assumes that BHs do not grow in mass during collisions with MS or giant stars \citep{Hurley2002,Breivik2020,Rodriguez2022}. BHs can grow in mass via collisions with white dwarfs, although these are incredibly rare. BHs, however, can grow via mass accretion from a binary companion.}

In \autoref{fig:mbh_hist2}, we show the distributions of $\mbh$ for all BHs formed via any channel inside the cluster during the whole simulated lifetime, $\tcluster/\gyr=13.8$. Note that we do not consider the changes that may come to the BHs after they are ejected from the cluster, for example, via mass transfer or GW-driven mergers outside the cluster. The $\mbh$ distributions are very similar for all models with widely different $\fbhighmod$. The only statistically significant difference can be seen at low $\log(\mbh/\msun)\lesssim0.5$; high $\fbhighmod$ models produce fewer low-mass BHs. This is a relic of the early evolution (\autoref{subs:earlytime}, \autoref{fig:nbh}). 

In all models, independent of $\fbhighmod$, the total number of BHs produced is dominated by stellar core-collapse; only $\sim1$--$2\%$ of all BHs form via BH-BH mergers (\autoref{fig:bhsall}, top). However, the picture changes completely when only high-mass BHs are considered. Production of high-mass BHs is dominated by BH-BH mergers for all $\fbhighmod$. Moreover, the number of high-mass BHs produced via BH-BH mergers do not show any significant correlation with $\fbhighmod$. Independent of $\fbhighmod$, the models create $\approx20$ high-mass BHs via BH-BH mergers per model compared to the maximum $\approx4$ via core-collapse of stellar collision products (\autoref{tab:ccbh}). As a result, at late times, the total number of high-mass BHs produced by a cluster overall do not show a correlation with $\fbhighmod$ (\autoref{fig:bhsall}, middle) despite the strong positive correlation between $\fbhighmod$ and high-mass BHs produced via core-collapse of stellar collision products (\autoref{fig:nbh}). Interestingly, even the most massive BH is typically formed via BH-BH mergers for all $\fbhighmod$ models except $\fbhighmod=1$ where $\mbhmax$ from core-collapse of stellar collision products win by a whisker (\autoref{fig:bhsall}, bottom; \autoref{tab:ccbh}).   

\autoref{fig:mbh_time} shows the mass and formation time for all new BHs in two sets of models, with $\fbhighmod=1$ (blue) and $0$ (orange). The dots (open circles) denote BHs formed via stellar core-collapse (BH-BH merger). The top axis shows the corresponding redshift ($z$) assuming a typical present-day age for globular clusters, $\tcluster/\gyr\sim10$ \citep[e.g.,][]{VandenBerg_2013} as reference. As expected, high-mass BH formation via stellar core-collapse and BH-BH mergers operate on very different timescales- $\tcluster/\myr\lesssim10$ is exclusive to stellar core-collapse whereas BH-BH mergers typically operate much later. This is because after remnant formation and the resulting expansion of the cluster, it takes time for the BHs to mass segregate to the core and partake in strong encounters to form new BBHs that can merge afterwards. Models with $\fbhighmod=0$ form BHs via stellar core-collapse almost exclusively below the upper mass-gap (only one high-mass BH formed in one of the models). In contrast, models with $\fbhighmod=1$ can form BHs in the upper mass gap and beyond more efficiently. At later times, high-mass BHs formed via BH-BH mergers have similar properties and formation times independent of $\fbhighmod$. Interestingly, although $\fbhighmod$ plays a crucial role during early evolution to form high-mass BHs more efficiently, we find that isolated evolution of the same high-mass binaries in one of our cluster models with $\fbhighmod=1$ could not create {\emph any} high-mass BHs (\autoref{app:isolated}). This highlights the importance of dynamical processes.

Another interesting trend is that the dispersion in $\mbh$ is significantly higher in the $\fbhighmod=1$ models via both channels for all times. The high dispersion stems from more stellar collisions in the $\fbhighmod=1$ models which creates a more diverse population of pre-collapse high-mass stars giving birth to a more diverse population of BHs. Higher $\fbhighmod$ also helps mix BHs with relatively more disparate masses in the core simply because it is possible to bring to the core relatively lower-mass BH progenitors as part of a binary via mass segregation. In contrast, for low $\fbhighmod$, at any given time, only the most massive retained population of BHs sink to the core and take part in BH dynamics. Thus, increasing $\fbhighmod$ naturally increases the diversity of BHs a cluster can form. However, this difference in the mass distribution of BH-BH merger remnants disappear after $\tcluster/\gyr=2$; by then all differences between models with different $\fbhighmod$ are dynamically erased (\autoref{fig:mbh_time}).

Note that mass-gap BHs may also form if a BH, born with mass below but close to the mass gap, accretes sufficient mass from a companion to push it into the mass gap. So far, we have not discussed those simply because these BHs are not high-mass BHs at the time of birth. Overall, we find that the number of high-mass BHs formed via mass transfer is $\sim20$ for models with $\fbhighmod=1$, comparable to the ones formed via BH-BH merger, and decreases with decreasing $\fbhighmod$ (\autoref{tab:ccbh}). Although significant in number, especially for high $\fbhighmod$ models, the mass for these high-mass BHs is typically just above the threshold used in our models, with median $\mbh/\msun\approx 46$ and $\mbhmax/\msun\lesssim52$. Necessarily in binaries, these BHs preferentially take part in BH dynamics due to larger interaction cross sections and production of merging BBHs. In \autoref{bhbhmergers}, we discuss how high-mass BHs influenced by mass transfer may play a role in the production of high-mass BH-BH mergers in clusters with high $\fbhighmod$.  

\subsection{Demographics of Merging Binary Black Holes}
\label{bhbhmergers}
\begin{table*}
    \centering
    \hspace{-2.5cm}\begin{tabular}{cccccccc}
    \hline
        \multirow{3}{.7cm}{$\fbhighmod$} & \multicolumn{7}{c}{Full Cluster Life}  \\ \cline{2-8}
         &  \multicolumn{2}{c}{All}  &  \multicolumn{2}{c}{High-mass} &  \multicolumn{3}{c}{High-mass BBH Formation Channel} \\ \cline{2-8}
         & In-cluster & Ejected &  In-cluster & Ejected & Core-collapse & BBH Merger & Mass Transfer \\
         \hline 
         \hline
        1 & $48^{+15}_{-6}$ & $72^{+12}_{-3}$ & $5^{+1}_{-1}$ & $10^{+2}_{-0}$ & 17 & 54 & 82 \\
        0.75 & $54^{+10}_{-6}$ & $70^{+6}_{-3}$ & $3^{+1}_{-1}$ & $8^{+2}_{-2}$ & 15 & 46 & 52 \\
        0.5 & $49^{+7}_{-7}$ & $64^{+3}_{-5}$ & $4^{+1}_{-2}$ & $7^{+1}_{-1}$ & 11 & 44 & 55 \\
        0.25 & $34^{+7}_{-1}$ & $54^{+4}_{-0}$ & $2^{+2}_{-1}$ & $4^{+2}_{-1}$ & 5 & 51 & 14 \\
        0 & $30^{+3}_{-4}$ & $40^{+5}_{-3}$ & $2^{+1}_{-1}$ & $4^{+1}_{-1}$ & 0 & 62 & 0\\
        \hline
        \hline
         & \multicolumn{7}{c}{$0<z<1$}\\ \cline{2-8}
        1 &  $10^{+2}_{-2}$ & $26^{+4}_{-3}$ & $0^{+0}_{-0}$ & $3^{+1}_{-1}$ & 4 & 16 & 15 \\
        0.75 & $12^{+1}_{-4}$ & $26^{+4}_{-3}$ & $0^{+0}_{-0}$ & $2^{+2}_{-1}$ & 3 & 15 & 7 \\
        0.5 & $12^{+2}_{-6}$ & $24^{+2}_{-2}$ & $0^{+0}_{-0}$ & $2^{+1}_{-1}$ & 2 & 14 & 7 \\
        0.25 & $8^{+3}_{-4}$ & $22^{+1}_{-2}$ & $0^{+0}_{-0}$ & $2^{+1}_{-1}$ & 0 & 17 & 4 \\
        0 & $10^{+5}_{-2}$ & $20^{+3}_{-3}$ & $0^{+1}_{-0}$ & $1^{+0}_{-0}$ & 0 & 14 & 0 \\
        \hline
    
    \end{tabular}
    \caption{The demographics of BBHs merging within a Hubble time in our models. For each $\fbhighmod$, we show the numbers for in-cluster mergers and mergers after ejection from the clusters overall and those involving high-mass BHs. Furthermore, we show the contributions from different formation channels to the high-mass merging BBHs. The top part of the table shows these numbers throughout the $\approx14\,\gyr$ lifetime of the models, whereas, the bottom part shows the corresponding numbers within $0<z<1$ (corresponds to a look-back time of $\approx 8.4\,\gyr$). All numbers with errors denote the medians and the quartiles obtained from the different realisations within models with a given $\fbhighmod$. The numbers without errorbars, given for formation channels of high-mass merging BBHs are total over these realisations due to their small per-model numbers. 
    }
    \label{tab:bbh}
\end{table*}

Merging BBHs are of special interest because they may be detected by the GWs they emit during mergers. We now focus our attention to the BBHs that merge within a Hubble time produced in our models. For this we take into account in-cluster mergers as well as those that merge via GW radiation after being ejected from the cluster \citep{Peters_1964}. \autoref{fig:bhsmergers} shows the individual pre-merger $\mbh$ distributions for merging BBHs. Interestingly, all models independent of $\fbhighmod$ can produce merging BBHs in the mass gap and beyond. However, there is a clear correlation between the number of merging BBHs involving high-mass BHs and $\fbhighmod$. For example, the median number of high-mass merging BBHs per model increases systematically from $7$ to $17$ as $\fbhighmod$ increases from $0$ to $1$ (\autoref{tab:bbh}).

\begin{figure}
    \centering
    \epsscale{1.1}
    \plotone{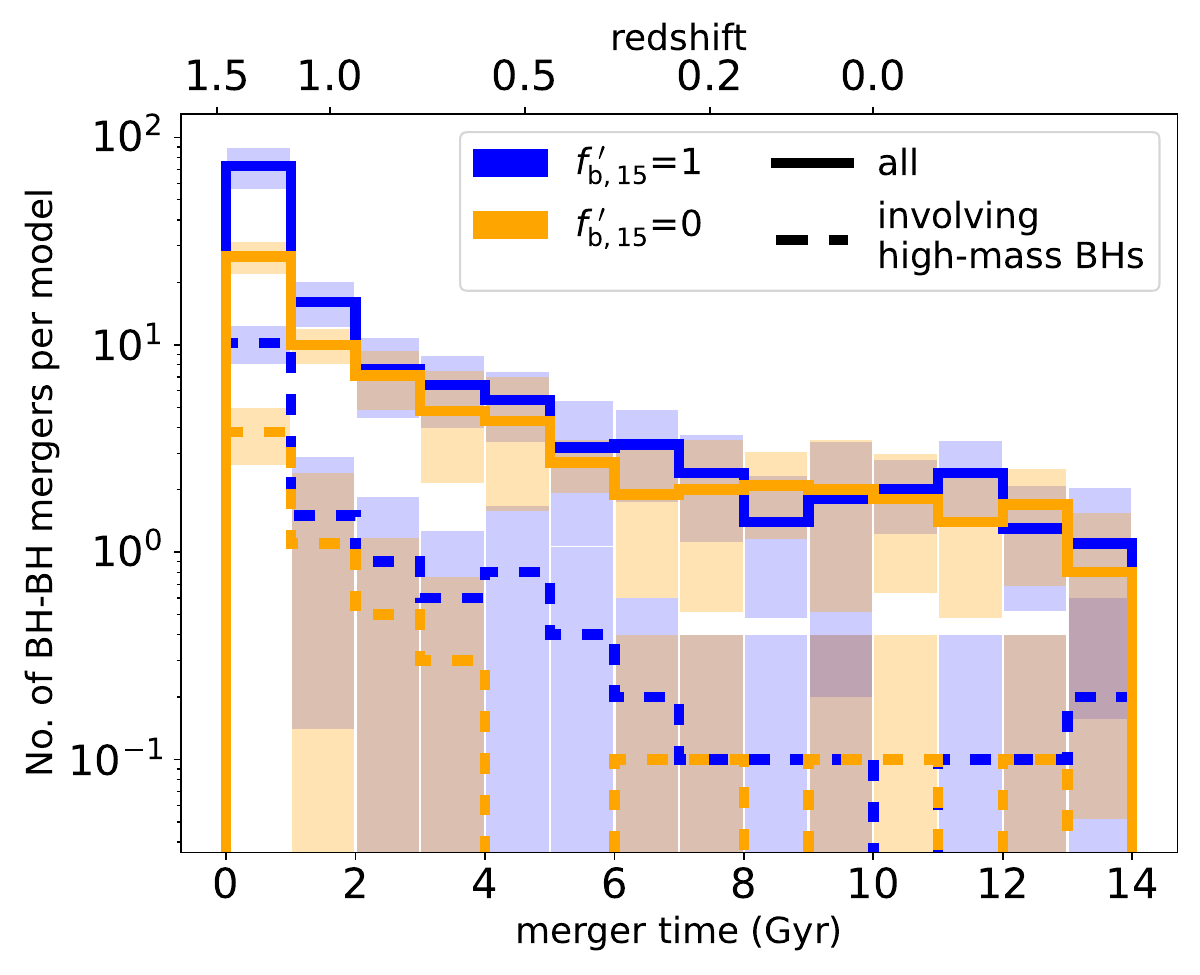}
    \caption{Histograms for the times of BBH mergers produced in our models, in-cluster as well as ejected, per cluster. The histograms and the shaded regions represent the mean and $1\sigma$ between realisations. Blue (orange) represents models with $\fbhighmod=1$ (0). Solid (dashed) line represents all BBH mergers (BBH mergers involving at least one high-mass BH). The top axis represents $z$ assuming $z=0$ is cosmic time $t_c=14\,\gyr$ and all model clusters formed $10\,\gyr$ ago.}
    \label{fig:bbhmr}
\end{figure}

Differences in the early production of high-mass BHs and their early mergers are primarily responsible for the positive correlation between $\fbhighmod$ and the number of merging high-mass BBHs. Moreover, as $\fbhighmod$ increases, so does the number of high-mass BHs formed via mass transfer (\autoref{tab:ccbh}). Although there is a very significant difference in both the total number of mergers as well as the number of high-mass mergers for merger time, $\tmerge/\gyr<1$, these differences reduce at later times (\autoref{fig:bbhmr}). In particular, after $\tcluster/\gyr\gtrsim4$, there is little statistical difference between the models with these vastly different $\fbhighmod$. 

\begin{figure}
    \centering
    \epsscale{1.1}
    \plotone{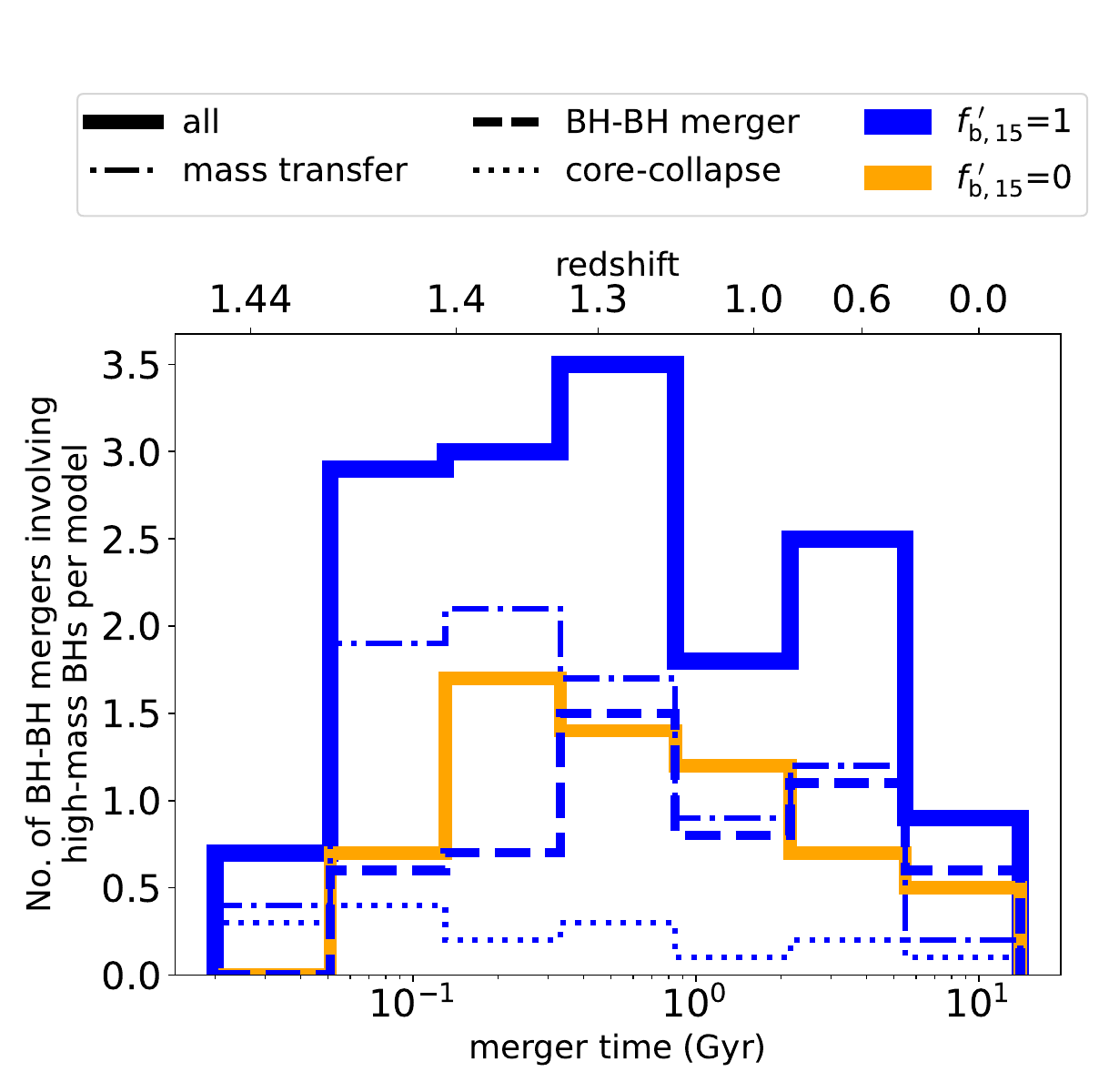}
    \caption{Histogram for the times of BBH mergers involving at least one high-mass BH produced per cluster. Blue (orange) represents $\fbhighmod=1$ (0). Dashed, dotted, dash-dotted, and solid histograms denote high-mass merging BHs produced via previous BH-BH mergers, core-collapse of stellar merger products, mass-transfer in a binary which pushes a normal BH into the upper mass gap, and all channels combined, respectively. All high-mass merging BHs form via previous BH-BH mergers in the models with $\fbhighmod=0$. In contrast, the high-mass merging BHs may be produced via all three channels in, e.g., the models with $\fbhighmod=1$ (also see \autoref{tab:bbh}). Merger time is calculated from $\tcluster=0$. The top axis denotes indicative $z$ assuming $z=0$ is cosmic time $t_c=14\,\gyr$ and all model clusters formed $10\,\gyr$ ago.}
    \label{fig:bbhmr2}
\end{figure}
The picture becomes clearer if the formation channel of the high-mass BH that is participating in the merger is noticed as a function of $\tmerge$ (\autoref{fig:bbhmr2}). In the models with $\fbhighmod=0$, all high-mass BHs in the merging BBHs come from previous BH-BH mergers that remain in cluster, dynamically acquire another BH companion, and undergo higher generation mergers. In contrast, there are multiple possibilities in the models with high $\fbhighmod$. Here, the high-mass BHs could be created from core-collapse of stellar merger products, previous BH-BH mergers, and also accretion of mass from a non-BH companion that pushes a normal BH into the upper mass gap (\autoref{tab:bbh}). For example, in case of the $\fbhighmod=1$ models, the contribution to high-mass merging BBHs formed via stellar core-collapse of merger products remains below the other channels for all $\tmerge$. Although the mass transfer channel can dominate the number of merging high-mass BHs in the $\fbhighmod=1$ models, the mass of these BHs remains close to the lower edge of the mass-gap irrespective of $\fbhighmod$. We find that the median mass for all {\em formally} mass-gap BHs formed via mass transfer is $46\,\msun$, just above the threshold used in our models, independent of $\fbhighmod$. Given the uncertainties in the exact boundaries of the mass gap, in reality, most high-mass BHs formed via mass-transfer may not be considered as such if their mergers are observed. After $\tmerge/\gyr\gtrsim6$, the dominant formation channel for the high-mass merging BBHs is previous BH-BH mergers. Interestingly, the numbers of high-mass BHs in merging BBHs produced via previous BH-BH mergers is insensitive to $\fbhighmod$ (\autoref{tab:bbh}).  

\begin{figure}
    \centering
    \epsscale{1.1}
    \plotone{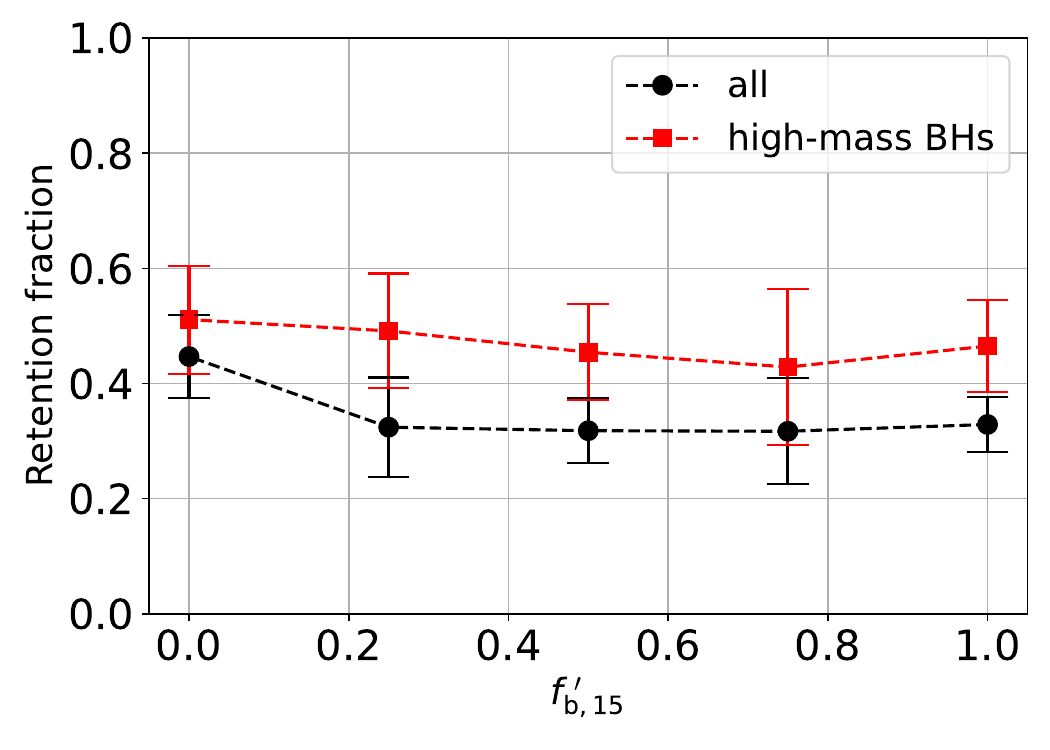}
    \caption{The retention fraction of BHs that form via BH-BH merger in our simulations as a function of $\fbhighmod$. Black (red) represents all (high-mass) BH-BH mergers. Markers (errorbars) denote the mean ($1\sigma$) of the retention fractions from different model realizations.}
    \label{fig:retfracr}
\end{figure}
It is important to note that the retention of a previous BH-BH merger product in the cluster is strongly dependent on the recoil from the GW emission. We take this into account (\autoref{sec:methods}). \autoref{fig:retfracr} shows the fraction of retained BH-BH merger products as a function of $\fbhighmod$. About $30-40\%$ of all BBH mergers are retained in our models. However, for high-mass BHs, expected to receive a relatively smaller recoil due to their larger mass, $\sim40-50\%$ are retained. Furthermore, the retention fraction of BBH merger products is independent of $\fbhighmod$.

\begin{figure}
    \centering
    \epsscale{1.1}
    \plotone{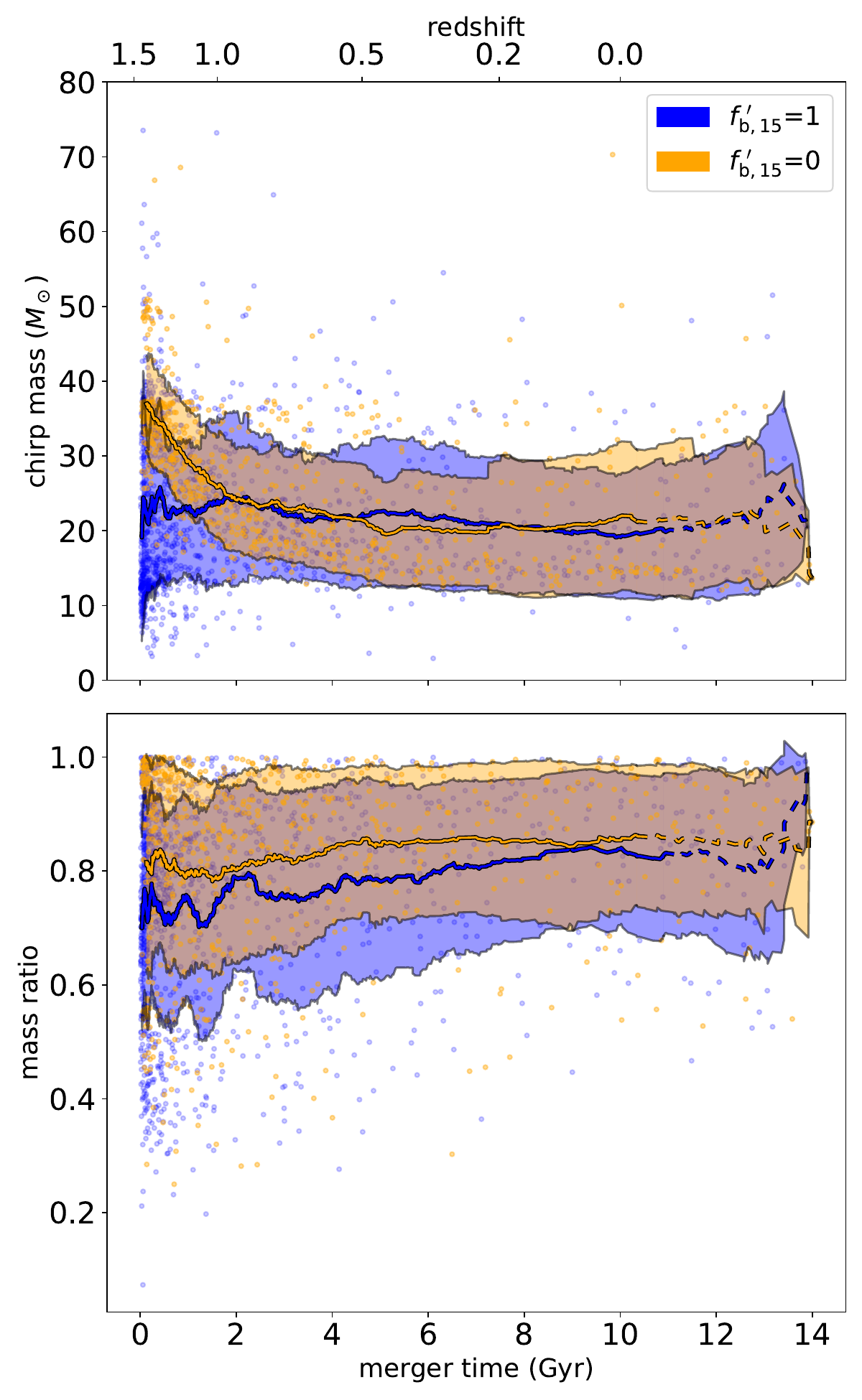}
    \caption{Chirp mass (\emph{top}) and mass ratio (\emph{bottom}) as a function of $\tmerge$ for all BBH mergers combining all realisations from models with $\fbhighmod=1$ (blue) and $0$ (orange). Lines with shaded regions represent the moving average and $1\sigma$ calculated using 101 chronologically sorted BBH mergers. The dashed parts of the lines near the two ends indicate that the moving average is calculated on a declining number of mergers. }
    \label{fig:chirp}
\end{figure}

\autoref{fig:chirp} shows the distributions for chirp mass, $\mchirp\equiv(\mbh_1 \mbh_2)^{3/5}/(\mbh_1 + \mbh_2)^{1/5}$ \citep[e.g.,][]{PhysRevD.49.2658} and mass ratio, $\qbh\equiv\mbh_1/\mbh_2$, where $\mbh_1<\mbh_2$ for merging BBHs as a function of $\tmerge$ for the models with $\fbhighmod=0$ and $1$. Clearly, $\mchirp$ for the merging BBHs show little difference for $\tmerge/\gyr\gtrsim2$ despite the difference in $\fbhighmod$. However, the models with $\fbhighmod=0$ show a systematically higher $\mchirp$ for earlier $\tmerge$. When none of the high-mass stars are in binaries, such as the models with $\fbhighmod=0$, the most massive single stars sink first to the core due to mass segregation. Because of this dynamical mass filtering, the BHs produced would only find other BHs of similar high mass to interact with within the core at any given time \citep[e.g.,][]{Breen_2013,Morscher_2013,Morscher_2015,Chatterjee_2017,Kremer_2019}. As a result, in the $\fbhighmod=0$ models, dynamics naturally pushes $\mchirp$ towards higher values. In contrast, if the cluster contains significant high-mass stellar binaries, such as our models with $\fbhighmod=1$, lower-mass stars as part of binaries can also sink to the core via mass segregation at the same time with significantly higher-mass stars. As a result, BHs with a much broader range in $\mbh$ may be able to interact in the core, dynamically form BBHs, and merge. Due to the same reason, the merging BBHs tend to have a higher $\qbh$ in the $\fbhighmod=0$ models compared to those in the models with $\fbhighmod=1$ (\autoref{fig:chirp}). Indeed, this relative difference in early mass segregation and the range in stellar masses that can arrive in the core between different $\fbhighmod$ models is the key to understanding a variety of our results. For example, the same process is responsible for the higher dispersion in $\mbh$ for models with $\fbhighmod=1$ at any given formation time $\lesssim2\,\gyr$ (\autoref{fig:mbh_time}). The same process is also responsible for a much wider spectrum of $\mbh$ in merging systems for models with $\fbhighmod>0$ compared to those found in the $\fbhighmod=0$ models (\autoref{fig:bhsmergers}). 

\section{Summary and conclusions}
\label{sec:conclusions}
Discovery of BBH mergers with pre-merger individual source-frame masses in or above the upper mass gap has generated wide-spread interest in thorough investigations for the formation channels of these BHs. \citetalias{Gonzalez2021} showed that a young, massive, and dense star cluster with $\fbhigh=1$ can produce BHs in the upper mass gap and beyond from core-collapse of high-mass stellar collision products. In contrast, they showed that similar models with $\fbhigh=0$ fail to produce the high-mass BHs. We have investigated this in more detail. In addition to the difference in the initial high-mass stellar binary fraction, the \citetalias{Gonzalez2021} models with $\fbhigh=0$ also contained $\approx74\%$
fewer high-mass stars compared to their models with $\fbhigh=1$. We carefully disentangle the effects of varying the fraction of high-mass stars in binaries and $\Nhigh$. For this, we create models with identical stellar populations, but vary the fraction of high-mass stars in binaries, $\fbhighmod=0$ to $1$ (\autoref{sec:methods}). Moreover, we study the long-term evolution of these models focusing on the overall formation of high-mass BHs from a variety of channels. We study the time-dependent relative importance for these channels and implications for merging systems as a function of look-back time or $z$. We summarise our key results below. 

\begin{itemize}
    \item BH formation via stellar core-collapse completes within $\tcluster/\myr\lesssim35$ in our models. Throughout $\tcluster/\myr<35$, most binary high-mass stars retain their primordial companions (\autoref{fig:time_frac}), and single high-mass stars remain single. Increasing $\fbhighmod$ allows faster mass segregation which brings more high-mass stars to the core before they could form remnants. Thus, despite containing identical $\Nhigh$, very different populations of stars arrive to the core via mass segregation in the models with varying $\fbhighmod$. This effectively alters the population of high-mass stars that can effectively collide/merge with other stars and form high-mass BHs via stellar core collapse. As a result, the number of high-mass BHs produced as well as $\mbhmax$ produced via stellar core-collapse show clear positive correlation with $\fbhighmod$ (\autoref{fig:mbh_hist}, \ref{fig:nbh}). 
    \item Most collisions/mergers take place within the cluster core (\autoref{fig:coll_time}). Although the high-mass stars spend most of their life before remnant formation with their primordial companions, most high-mass stellar collisions happen between non-primordial companions (e.g., $\approx97\%$ for models with $\fbhighmod=1$; \autoref{subs:earlytime}). Collisions/mergers are also temporally concentrated between $3\lesssim\tcluster/\myr\lesssim5$ independent of $\fbhighmod$. This is because of the race between two-body relaxation and mass segregation, which increase $\rho_c$, and high-mass stellar evolution and remnant formation, which rapidly expand the cluster via mass loss and decrease $\rho_c$. As a result, $\rho_c$ peaks during this time window. Furthermore, immediately before remnant formation, the stars achieve their largest physical sizes enhancing collision/merger potential. 
    \item At later times, production of high-mass BHs is dominated by BH-BH mergers (\autoref{fig:mbh_hist2}, \autoref{fig:mbh_time}). The number and $\mbhmax$ for high-mass BHs produced via BH-BH mergers show no correlation with $\fbhighmod$ (\autoref{fig:bhsall}). As a result, the overall number and $\mbhmax$ produced via all channels combined throughout the entire history of a cluster remains insensitive to $\fbhighmod$. 
    \item Increasing $\fbhighmod$ increases the diversity of stars that can collide/merge during the early times of a cluster's life. This also increases the diversity of BHs that can interact with each other resulting in a wider spread in BH masses a cluster can produce (\autoref{fig:mbh_time}). Pre-merger individual $\mbh$ for merging BBHs show a larger spread as $\fbhighmod$ increases due to the same reason (\autoref{fig:bhsmergers}).  
    \item The overall number of high-mass BBH mergers show a clear correlation with $\fbhighmod$ (\autoref{fig:bhsmergers}). However, this is due to the differences in the number of merging BBHs with $\tmerge/\gyr<1$. The number of BBH mergers, both high-mass and overall, at $\tmerge/\gyr\gtrsim2$ shows no statistical difference between models with varying $\fbhighmod$ (\autoref{fig:bbhmr}). Assuming a typical present-day age of $10\,\gyr$ for these clusters as a reference, identification of the effects of $\fbhighmod$ on the total produced number of merging BBHs would require sensitivity beyond $z=1$. 
    \item The mean $\mchirp$ for merging BBHs produced in $\fbhighmod=0$ models is significantly higher compared to those produced in $\fbhighmod=1$ models for $\tmerge/\gyr\lesssim1$. The models with $\fbhighmod$ in between $0$ and $1$ show a systematic decrease in the mean $\mchirp$ with increasing $\fbhighmod$. However, these differences disappear for $\tmerge/\gyr>2$ (\autoref{fig:chirp}). The mean $\qbh$ for merging BBHs produced in the $\fbhighmod=0$ models tend to be higher compared to those produced in the $\fbhighmod=1$ models. This is simply because higher $\fbhighmod$ facilitates production of and interactions between BHs with a wider spectrum in $\mbh$ (\autoref{fig:bhsmergers}), especially during the early evolution of a cluster.  
\end{itemize}

In this study, we perform controlled experiments to clearly identify the effects of $\fbhighmod$ on the production of high-mass BHs in dense star clusters. We find that increasing $\fbhighmod$ enhances the rate of high-mass stellar collisions/mergers enabling the production of high-mass BHs via core-collapse of merger products. However, we find that the effects are nuanced and the rich physics in star clusters produce trends that are seemingly contradictory. For example, although the production of high-mass BHs via core-collapse of high-mass collision products is strongly correlated with $\fbhighmod$, the overall production of high-mass BHs is dominated by BH-BH mergers over the full lifetime of a cluster. Since BH-BH mergers show no correlations with $\fbhighmod$, the overall number of high-mass BHs a cluster can produce show no significant correlation with $\fbhighmod$. While during early evolution, most high-mass stars remain bound to their primordial companions or remain single, only $\approx3\%$ of high-mass collisions happen between primordial companions. While the overall number of merging BBHs a cluster produces show a positive correlation with $\fbhighmod$, this effect stays limited to $\tmerge/\gyr
\lesssim1$; varying $\fbhighmod$ shows no effect in the number of merging BBHs for any $\tmerge/\gyr\gtrsim2$. While higher $\fbhighmod$ models produce higher numbers of high-mass BHs and higher $\mbhmax$ during the early evolution of the cluster, the mean $\mchirp$ as well as $\qbh$ are lower for any $\tmerge/\gyr\lesssim1$. These results highlight the richness in the interrelated physical processes and outcomes star cluster dynamics studies can reveal. We also highlight the importance of expanding the discovery horizon to higher redshifts by the existing and planned GW detectors in order to probe the rich and interesting early evolution of massive and dense star clusters expected to have formed during an epoch of heightened star formation rate in the Universe.       
\\

% \begin{acknowledgments}

We thank the anonymous referee for thoughtful questions and comments. AK acknowledges support from TIFR's graduate fellowship. SC acknowledges support from the Department of Atomic Energy, Government of India, under project no. 12-R\&D-TFR-5.02-0200 and RTI 4002. All simulations were performed on a cloud-based high-performance computing solution on Azure.

% \end{acknowledgments}

\software{\texttt{CMC} \citep{Rodriguez2022}, \texttt{COSMIC} \citep{Breivik2020}, \texttt{cmctoolkit} \citep{cmctoolkit}, \texttt{Anaconda} \citep{anaconda}, \texttt{NumPy} \citep{harris2020array}, \texttt{pandas} \citep{reback2020pandas,mckinney-proc-scipy-2010}, \texttt{matplotlib} \citep{Hunter:2007}, \texttt{Astropy} \citep{astropy:2022}.}

% \bibliography{Mass_gap_BHs,extra_ref}{}
\bibliographystyle{aasjournal}

\appendix
\restartappendixnumbering

\section{Agreement with direct $N$-body models}
\label{app:dnb}
\begin{figure}
    \centering
    \plotone{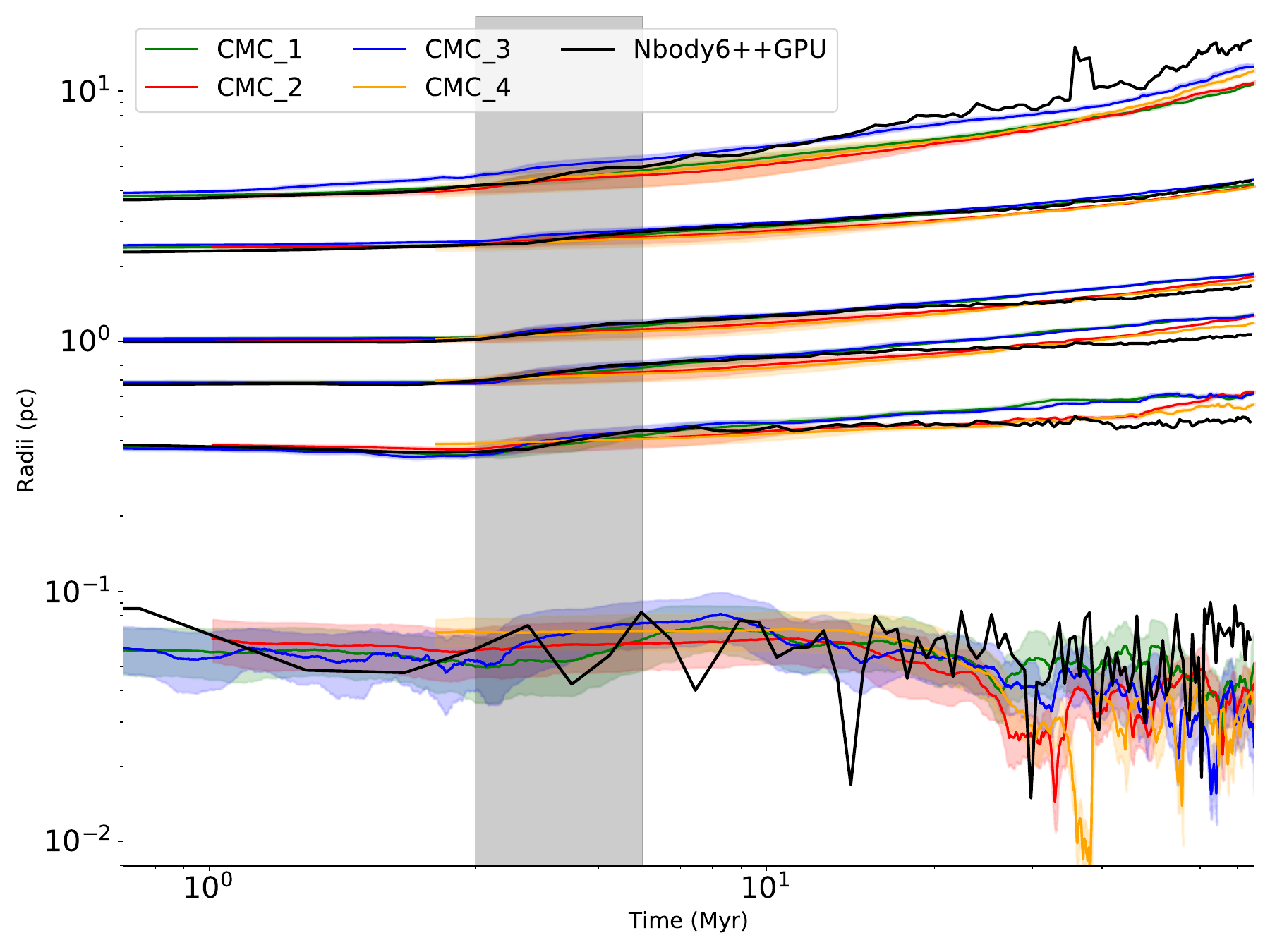}
    \caption{The evolution of the (0.1, 10, 30, 50, 90, 99)\% Lagrange radii of a model star cluster with $10^5$ initial objects and $\fb=5\%$. The black curves are from the \texttt{Nbody6++GPU} run, while the colored curves with $1-\sigma$ fluctuations are from four different model realisations simulated using $\cmc$. The vertical shaded region shows the time window where most of the high-mass star collisions happen in our simulations (\autoref{fig:coll_time}).}
    \label{fig:app_dnb}
\end{figure}

The effects of $\fbhighmod$ on the production of high-mass BHs in a star cluster, especially during early evolution, is closely related to the faster mass segregation of high-mass stellar binaries by a factor of $\sim(1+\qhigh)$. Hence, we test the agreement between star cluster models created using \cmc\ and \texttt{Nbody6++GPU} \citep{Wang_2015,Kamlah_2022}, one of the most advanced publicly available direct $N$-body codes. We particularly focus on the agreement in the distribution of mass within a star cluster during the early evolution. We simulate a realistic star cluster model with $10^5$ initial objects (singles/binaries) and $\fb=5\%$ for $\sim70\,\myr$ using both codes. \autoref{fig:app_dnb} shows the evolution of various Lagrange radii for the cluster models created using these codes. The vertical gray shaded region denotes the time window when most of the high-mass collisions take place in our models (\autoref{fig:coll_time}). We observe that all the Lagrange radii, thus the mass distribution, match very well between the two codes. This result strengthens the reliability of our $\cmc$ models to properly capture the effects of mass segregation during this early phase of evolution.

\section{Comparison with isolated binary evolution}
\label{app:isolated}

\begin{figure}
    \centering
    \plotone{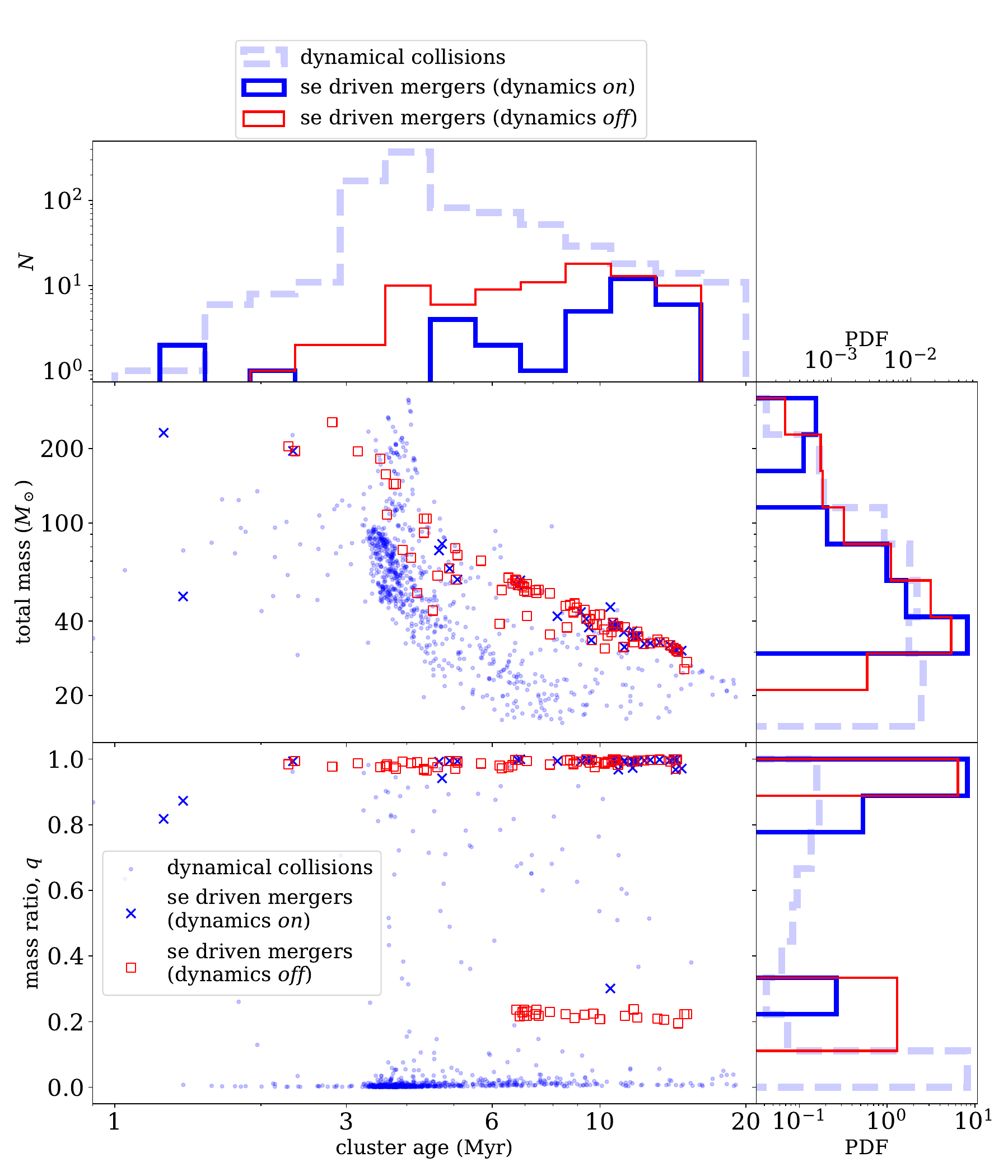}
    \caption{Comparison of all high-mass collisions/mergers in our simulations with (blue) and without (red) dynamics. The bottom panel shows the mass ratio of the colliding/merging members ($M_1/M_2$, where $M_1<M_2$) as a function of the time of the collision/merger. On occasion, a dynamical collision could take place between more than two (three or four) stars. In that case, we show the mass ratio of the two most massive colliding members. The middle panel shows the total mass of all the colliding/merging members along the $y$-axis. In both of these panels, dots represent dynamical collisions, while crosses and squares represent stellar evolution driven mergers with and without dynamics, respectively. By definition, the simulation without dynamics doesn’t host any collisions. The upper and side panels show the corresponding distributions of the time, total mass, and mass ratio. In these panels, the dashed-faded-blue histograms represent collisions, while the solid blue and red histograms represent stellar evolution driven mergers with and without dynamics, respectively.}
    \label{fig:app_collisions}
\end{figure}

\begin{figure*}
    \centering
    \plotone{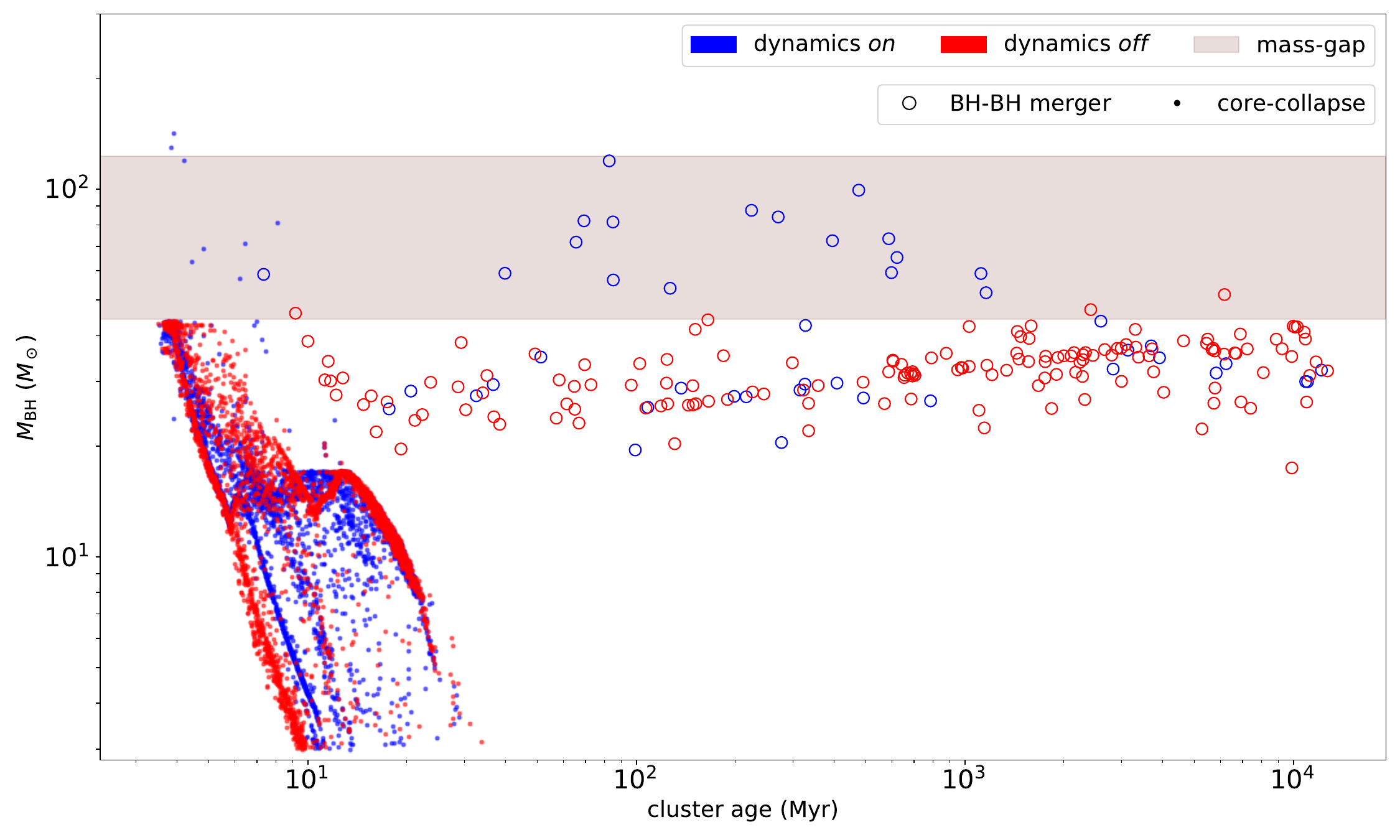}
    \caption{The same as \autoref{fig:mbh_time}, except here we compare BH formation in models with (blue) and without (red) dynamics.}
    \label{fig:app_BH}
\end{figure*}

High-mass stellar collisions during the early evolution of a star cluster play a crucial role in the creation of high-mass BHs (e.g., \autoref{subs:earlytime}). We have found in the main text that these collisions are almost always dynamically produced, typically during binary-mediated strong encounters. High-mass stellar binaries may also merge simply due to binary stellar evolution in isolation without getting affected by dynamics. It is thus interesting to check whether the same high-mass binaries in our star cluster models would merge in isolation to produce the progenitors of high-mass BHs. We have simulated one of our model initial clusters with $\fbhighmod=1$ by turning off \emph{all} dynamical processes. This essentially captures the fate of the same stars and binaries if allowed to evolve in isolation. In \autoref{fig:app_collisions}, we compare the collisions/mergers between these two cases, with (blue) and without dynamics (red). We find that the prominent peak in collisions/mergers between $3-5\,\myr$ (\autoref{fig:coll_time}) in our simulations is absent if the same binaries are evolved in isolation. In fact, with or without dynamics, a small fraction of all high-mass binaries would merge through binary stellar evolution, but these mergers are roughly uniformly distributed over time. The peak observed in \autoref{fig:coll_time} comes from the increase of $\rho_c$ due to relaxation and mass segregation until remnant formation begins which rapidly decreases $\rho_c$. Although we find low numbers of mergers, it appears that with dynamics the same binaries may merge via stellar evolution somewhat less often compared to in isolation. This may be because dynamical processes can often significantly change these high-mass binaries via exchange interactions.

Interestingly, the mass ratios for dynamical collisions and stellar-evolution driven mergers show very different distributions. Whereas, the colliding stars typically have $q\lesssim0.1$, mergers happen with typically $q\sim1$. In isolation, in addition to the dominant peak near $q\sim1$, we also find a smaller peak near $q\sim0.2$.  
The peak near $q\sim0.2$ is entirely created by mergers involving at least one naked He star. Based on the prescriptions adopted in \cosmic\  \citep{Breivik2020}, we have found cases where a naked He star merges with a higher-mass MS star to form a BH or a NS. Alternatively, it can also merge with a higher-mass core He burning star to form a new core He burning star.  
We also find that the total mass of mergers/collisions decreases with time in both simulations. This is expected as higher-mass stars live shorter and also sink quicker to the cluster center to promote collisions (when stellar dynamics is active) compared to lower-mass stars.

\autoref{fig:app_BH} compares the mass and formation time of BHs in the simulations with and without dynamics. It is clear that high-mass BHs are not created with a mass well within or above the upper mass-gap without dynamics. We find only a few cases of (barely) mass-gap BHs produced via BH-BH mergers at later times. Clearly, formation of high-mass BHs is critically dependent on the dynamical processes active in star clusters.

\end{document}